\documentclass[aps,prd,reprint,twocolumn,superscriptaddress,preprintnumbers,nofootinbib]{revtex4-2}
\usepackage[english]{babel}
\usepackage[utf8]{inputenc}
\usepackage[colorinlistoftodos, color=green!40, prependcaption]{todonotes}
\usepackage{amsthm}
\usepackage{mathtools}
\usepackage{physics}
\usepackage{xcolor}
\usepackage{graphicx}
\usepackage[left=23mm,right=13mm,top=35mm,columnsep=15pt]{geometry} 
\usepackage{adjustbox}
\usepackage{placeins}
\usepackage[T1]{fontenc}
\usepackage{lipsum}
\usepackage{csquotes}

\usepackage{filecontents,catchfile}
\usepackage[pdftex, pdftitle={Article}, pdfauthor={Author},colorlinks = true,allcolors = blue]{hyperref}

\newcommand\GiBUU{\textsc{GiBUU}}
\newcommand\eHIJING{e\textsc{HIJING}}
\newcommand\BeAGLE{\textsc{B}e\textsc{AGLE}}

\usepackage{lineno}

\begin{document}
\title{Dihadron Azimuthal Correlations in Deep-Inelastic Scattering Off Nuclear Targets}

\newcommand*{\ANL}{Argonne National Laboratory, Argonne, Illinois 60439}
\newcommand*{\ANLindex}{1}
\affiliation{\ANL}
\newcommand*{\ASU}{Arizona State University, Tempe, Arizona 85287-1504}
\newcommand*{\ASUindex}{2}
\affiliation{\ASU}
\newcommand*{\CSUDH}{California State University, Dominguez Hills, Carson, CA 90747}
\newcommand*{\CSUDHindex}{3}
\affiliation{\CSUDH}
\newcommand*{\UCR}{University of California Riverside, Riverside, California 92521}
\newcommand*{\UCRindex}{4}
\affiliation{\UCR}
\newcommand*{\CANISIUS}{Canisius College, Buffalo, NY}
\newcommand*{\CANISIUSindex}{5}
\affiliation{\CANISIUS}
\newcommand*{\CMU}{Carnegie Mellon University, Pittsburgh, Pennsylvania 15213}
\newcommand*{\CMUindex}{6}
\affiliation{\CMU}
\newcommand*{\CUA}{Catholic University of America, Washington, D.C. 20064}
\newcommand*{\CUAindex}{7}
\affiliation{\CUA}
\newcommand*{\SACLAY}{IRFU, CEA, Universit\'{e} Paris-Saclay, F-91191 Gif-sur-Yvette, France}
\newcommand*{\SACLAYindex}{8}
\affiliation{\SACLAY}
\newcommand*{\CNU}{Christopher Newport University, Newport News, Virginia 23606}
\newcommand*{\CNUindex}{9}
\affiliation{\CNU}
\newcommand*{\UCONN}{University of Connecticut, Storrs, Connecticut 06269}
\newcommand*{\UCONNindex}{10}
\affiliation{\UCONN}
\newcommand*{\DUQUESNE}{Duquesne University, 600 Forbes Avenue, Pittsburgh, PA 15282 }
\newcommand*{\DUQUESNEindex}{11}
\affiliation{\DUQUESNE}
\newcommand*{\FU}{Fairfield University, Fairfield CT 06824}
\newcommand*{\FUindex}{12}
\affiliation{\FU}
\newcommand*{\FERRARAU}{Universit\`{a} di Ferrara , 44121 Ferrara, Italy}
\newcommand*{\FERRARAUindex}{13}
\affiliation{\FERRARAU}
\newcommand*{\FIU}{Florida International University, Miami, Florida 33199}
\newcommand*{\FIUindex}{14}
\affiliation{\FIU}
\newcommand*{\FSU}{Florida State University, Tallahassee, Florida 32306}
\newcommand*{\FSUindex}{15}
\affiliation{\FSU}
\newcommand*{\GWUI}{The George Washington University, Washington, DC 20052}
\newcommand*{\GWUIindex}{16}
\affiliation{\GWUI}
\newcommand*{\INFNC}{INFN, Sezione di Catania, Via S. Sofia 64, 95123 Catania, Italy}
\newcommand*{\INFNCindex}{17}
\affiliation{\INFNC}
\newcommand*{\INFNFE}{INFN, Sezione di Ferrara, 44100 Ferrara, Italy}
\newcommand*{\INFNFEindex}{18}
\affiliation{\INFNFE}
\newcommand*{\INFNFR}{INFN, Laboratori Nazionali di Frascati, 00044 Frascati, Italy}
\newcommand*{\INFNFRindex}{19}
\affiliation{\INFNFR}
\newcommand*{\INFNGE}{INFN, Sezione di Genova, 16146 Genova, Italy}
\newcommand*{\INFNGEindex}{20}
\affiliation{\INFNGE}
\newcommand*{\INFNRO}{INFN, Sezione di Roma Tor Vergata, 00133 Rome, Italy}
\newcommand*{\INFNROindex}{21}
\affiliation{\INFNRO}
\newcommand*{\INFNTUR}{INFN, Sezione di Torino, 10125 Torino, Italy}
\newcommand*{\INFNTURindex}{22}
\affiliation{\INFNTUR}
\newcommand*{\INFNPAV}{INFN, Sezione di Pavia, 27100 Pavia, Italy}
\newcommand*{\INFNPAVindex}{23}
\affiliation{\INFNPAV}
\newcommand*{\ORSAY}{Universit\'{e} Paris-Saclay, CNRS/IN2P3, IJCLab, 91405 Orsay, France}
\newcommand*{\ORSAYindex}{24}
\affiliation{\ORSAY}
\newcommand*{\JMU}{James Madison University, Harrisonburg, Virginia 22807}
\newcommand*{\JMUindex}{25}
\affiliation{\JMU}
\newcommand*{\KNU}{Kyungpook National University, Daegu 41566, Republic of Korea}
\newcommand*{\KNUindex}{26}
\affiliation{\KNU}
\newcommand*{\LAMAR}{Lamar University, 4400 MLK Blvd, PO Box 10046, Beaumont, Texas 77710}
\newcommand*{\LAMARindex}{27}
\affiliation{\LAMAR}
\newcommand*{\UMICH}{University of Michigan, 500 S State St, Ann Arbor, MI 48109}
\newcommand*{\UMICHindex}{28}
\affiliation{\UMICH}
\newcommand*{\MISS}{Mississippi State University, Mississippi State, MS 39762-5167}
\newcommand*{\MISSindex}{28}
\affiliation{\MISS}
\newcommand*{\UNH}{University of New Hampshire, Durham, New Hampshire 03824-3568}
\newcommand*{\UNHindex}{29}
\affiliation{\UNH}
\newcommand*{\NMSU}{New Mexico State University, PO Box 30001, Las Cruces, NM 88003, USA}
\newcommand*{\NMSUindex}{30}
\affiliation{\NMSU}
\newcommand*{\NSU}{Norfolk State University, Norfolk, Virginia 23504}
\newcommand*{\NSUindex}{31}
\affiliation{\NSU}
\newcommand*{\OHIOU}{Ohio University, Athens, Ohio  45701}
\newcommand*{\OHIOUindex}{32}
\affiliation{\OHIOU}
\newcommand*{\ODU}{Old Dominion University, Norfolk, Virginia 23529}
\newcommand*{\ODUindex}{33}
\affiliation{\ODU}
\newcommand*{\JLUGiessen}{II Physikalisches Institut der Universitaet Giessen, 35392 Giessen, Germany}
\newcommand*{\JLUGiessenindex}{34}
\affiliation{\JLUGiessen}
\newcommand*{\RPI}{Rensselaer Polytechnic Institute, Troy, New York 12180-3590}
\newcommand*{\RPIindex}{35}
\affiliation{\RPI}
\newcommand*{\URICH}{University of Richmond, Richmond, Virginia 23173}
\newcommand*{\URICHindex}{36}
\affiliation{\URICH}
\newcommand*{\ROMAII}{Universit\`{a} di Roma Tor Vergata, 00133 Rome Italy}
\newcommand*{\ROMAIIindex}{37}
\affiliation{\ROMAII}
\newcommand*{\MSU}{Skobeltsyn Institute of Nuclear Physics, Lomonosov Moscow State University, 119234 Moscow, Russia}
\newcommand*{\MSUindex}{38}
\affiliation{\MSU}
\newcommand*{\SCAROLINA}{University of South Carolina, Columbia, South Carolina 29208}
\newcommand*{\SCAROLINAindex}{39}
\affiliation{\SCAROLINA}
\newcommand*{\TEMPLE}{Temple University,  Philadelphia, PA 19122 }
\newcommand*{\TEMPLEindex}{40}
\affiliation{\TEMPLE}
\newcommand*{\JLAB}{Thomas Jefferson National Accelerator Facility, Newport News, Virginia 23606}
\newcommand*{\JLABindex}{41}
\affiliation{\JLAB}
\newcommand*{\UTFSM}{Universidad T\'{e}cnica Federico Santa Mar\'{i}a, Casilla 110-V Valpara\'{i}so, Chile}
\newcommand*{\UTFSMindex}{42}
\affiliation{\UTFSM}
\newcommand*{\BRESCIA}{Universit\`{a} degli Studi di Brescia, 25123 Brescia, Italy}
\newcommand*{\BRESCIAindex}{43}
\affiliation{\BRESCIA}
\newcommand*{\GLASGOW}{University of Glasgow, Glasgow G12 8QQ, United Kingdom}
\newcommand*{\GLASGOWindex}{44}
\affiliation{\GLASGOW}
\newcommand*{\YORK}{University of York, York YO10 5DD, United Kingdom}
\newcommand*{\YORKindex}{45}
\affiliation{\YORK}
\newcommand*{\VT}{Virginia Tech, Blacksburg, Virginia   24061-0435}
\newcommand*{\VTindex}{46}
\affiliation{\VT}

\newcommand*{\VIRGINIA}{University of Virginia, Charlottesville, Virginia 22901}
\newcommand*{\VIRGINIAindex}{47}
\affiliation{\VIRGINIA}
\newcommand*{\YEREVAN}{Yerevan Physics Institute, 375036 Yerevan, Armenia}
\newcommand*{\YEREVANindex}{48}
\affiliation{\YEREVAN}

\newcommand*{\NOWJLAB}{Thomas Jefferson National Accelerator Facility, Newport News, Virginia 23606}
\newcommand*{\NOWISU}{Idaho State University, Pocatello, Idaho 83209}
\newcommand*{\NOWKSU}{King Saud University, Riyadh, Saudi Arabia}

\author{S.J.~Paul}
\affiliation{\UCR}
\author{S.~Mor\'an}
\affiliation{\UCR}
\author{M.~Arratia}
\email[Corresponding author: ]{miguel.arratia@ucr.edu}
\affiliation{\UCR}
\author{W.K.~Brooks} 
\affiliation{\UTFSM}
\affiliation{\JLAB}
\author {H.~Hakobyan} 
\affiliation{\UTFSM}
\author {A.~El~Alaoui} 
\affiliation{\UTFSM}
\author {P.~Achenbach} 
\affiliation{\JLAB}
\author {J.S.~Alvarado} 
\affiliation{\ORSAY}
\author {W.R.~Armstrong} 
\affiliation{\ANL}
\author {H.~Atac} 
\affiliation{\TEMPLE}
% ...

\collaboration{The CLAS Collaboration}
\noaffiliation

\date{\today} 

\begin{abstract}
We measured the nuclear dependence of the di-pion azimuthal correlation function in deep-inelastic scattering (DIS) using the CEBAF Large Acceptance Spectrometer (CLAS) and a 5 GeV electron beam. As the nuclear-target size increases, transitioning from deuterium to carbon, iron, and lead, the correlation function broadens monotonically. Its shape exhibits a significant dependence on kinematics, including the transverse momentum of the pions and the difference in their rapidity. 
None of the various Monte-Carlo event generators we evaluated could fully replicate the observed correlation functions and nuclear effects throughout the entire phase space. As the first study of its kind in DIS experiments, this research provides an important baseline for enhancing our understanding of the interplay between the nuclear medium and the hadronization process in these reactions.

\end{abstract}

\maketitle

\section{Introduction} \label{sec:outline}

Understanding the complex dynamics of hadron production remains a frontier in the study of quantum chromodynamics (QCD) within nuclei. At the heart of this understanding are two crucial questions, as outlined in Ref.~\cite{LRP2023}. First, how are different hadrons produced in a single scattering event correlated with each other? Second, how does the process of hadronization manifest within a nucleus? To address these questions, detailed studies are crucial for understanding the relative importance of different effects and for determining key parameters, including hadron formation time, transport properties of nuclei, and the spatial and momentum distributions of quarks and gluons in the nucleus~\cite{Accardi:2009qv}. 

The study of hadron production in electron scattering off nuclei was notably advanced by HERMES measurements~\cite{HERMES:2000ytc,HERMES:2003icw,HERMES:2007plz,HERMES:2009uge,HERMES:2011qjb}, which have been expanded by CLAS measurements~\cite{CLAS:2011oae,CLAS:2021jhm,CLAS:2022oux,CLAS:2022asf}. However, these single-hadron measurements lack sufficient constraining power to clearly differentiate among various models, which encompass concepts such as gluon bremsstrahlung, ``pre-hadron'' states, intranuclear re-scattering of hadrons, absorption, or combinations thereof~\cite{Accardi:2009qv}. 

Di-hadron measurements have led to significant advancements, introducing new kinematic variables such as rapidity difference, which can help constrain the space-time structure of hadronization. The HERMES experiment first reported a comparison of the yields of hadron pairs produced in scattering off nuclei with those off deuterium~\cite{HERMES:2005mar}. Our recent results~\cite{CLAS:2022asf} have expanded upon these findings, revealing a significant nuclear dependence of the azimuthal separation of the pion pair.

\begin{figure}[h]
\centering
\includegraphics[width=0.98\columnwidth]{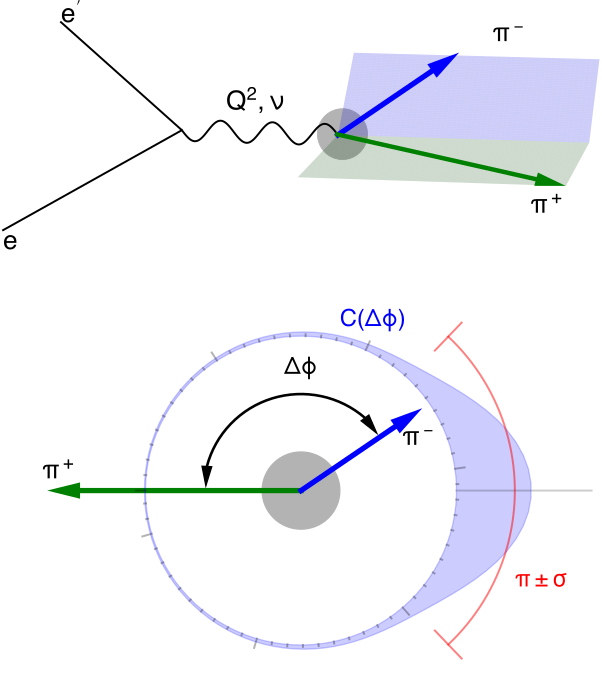}
\caption{Diagram illustrating di-pion production in a nucleus.  Top: side-view, illustrating that the momentum-transfer axis is used to define the azimuthal separation, $\Delta\phi$.  Bottom:  same reaction, as viewed from the direction of the momentum transfer.  The shaded area  represents the correlation function as a distribution in $\Delta\phi$, and the RMS width, $\sigma$, of this distribution (red).}
\label{fig:fig1}
\end{figure}

In this work, we extend our previous research by introducing, for the first time in the realm of DIS experiments, di-hadron angular correlation function measurements. Such measurements have been widely used in hadro-production experiments for the exploration of both cold and hot QCD matter, \textit{e.g.}, Refs.~\cite{STAR:2021fgw,LHCb:2023png,CMS:2012qk,CMS:2015fgy}. 
The di-pion correlation function is defined as:
\begin{equation}
    C(\Delta\phi) = C_{0}\times\frac{1}{N_{e'\pi^+}} \frac{dN_{e'\pi^+\pi^-}(\Delta\phi)}{d\Delta\phi},
    \label{eq:defC}
\end{equation}
where $N_{e'\pi^+}$ is the number of DIS events with a $\pi^+$ in the final state, $N_{e'\pi^+\pi^-}(\Delta\phi)$ refers to the number of $\pi^+\pi^-$ pairs that have an azimuthal-angle separation $\Delta\phi$ (that is, the angle between the plane containing the direction of one of the hadrons and that of the virtual photon, and the plane containing the other hadron's direction and the virtual photon's direction. 
See Fig.~\ref{fig:fig1}), and $C_{0}$ is a normalization factor, which is defined as: 
\begin{equation}
   C_{0}=  \frac{N^D_{e'\pi^+}}{ N^{D,\rm tot}_{e'\pi^+\pi^-}},
\end{equation}
where the superscript $D$ indicates that this value is being evaluated for the deuterium target. The same normalization factor, $C_0$, is applied to the nuclear data as well.  The superscript ``tot'' in the denominator indicates that this is integrated over the full range in $\Delta\phi$.  

We determine $C(\Delta\phi)$ across intervals of three variables, starting with the rapidity difference between the two pions, $\Delta Y = Y_1 - Y_2$, where rapidity is defined as:
\begin{equation}
    Y_h = \frac{1}{2}\ln \frac{E_h+p_{z,h}}{E_h-p_{z,h}}.
\end{equation}
Here, $p_{z,h}$ denotes the hadron momentum along the $\gamma^*$ direction and $E_h$ its energy. The other binning variables are the transverse momentum of the leading pion, $p^T_1$, and the sub-leading pion, $p^T_2$, relative to $\gamma^*$. We chose these three variables in particular because they are independent of boosts and rotations along the $\gamma^*$ direction.  

Additionally, we present three types of quantities derived from the correlation functions, namely the nuclear-to-deuterium ratio of the correlation functions, $R(\Delta\phi)=C_A(\Delta\phi)/C_D(\Delta\phi)$ (where $C_A$ and $C_D$ are the correlation functions for a nuclear target and deuterium, respectively), the RMS width, $\sigma$, of $C(\Delta\phi)$ around $\Delta\phi=\pi$, and the broadening, $b$, which is defined as:
\begin{equation}
    b=\sqrt{\sigma_A^2-\sigma_D^2},
    \label{eq:broadening}
\end{equation}
where $\sigma_A$ and $\sigma_D$ represent the RMS widths of the correlation functions for the nuclear and deuterium targets.   
This variable is anticipated to be sensitive to multiple scattering within nuclei, serving as an important probe for a variety of effects~\cite{Liang:2008vz,Xing:2012ii,Schafer:2013mza,Mueller:2016xoc,Cougoulic:2017ust,Alrashed:2021csd,Dupre:2023bnt}. 

We detail the experimental setup and event selection in Section~\ref{sec:setup}, discuss the systematic uncertainties in Section~\ref{sec:systematics}, present the results in Section~\ref{sec:Results}, and conclude in Section~\ref{sec:conclusions}.

\section{Experimental Method} \label{sec:setup}
This study follows methods used in Ref.~\cite{CLAS:2021jhm,CLAS:2022asf}, using the same dataset collected with the CLAS detector~\cite{CLAS:2003umf} and a 5.014~GeV electron beam. A dual-target system~\cite{Hakobyan:2008zz} simultaneously exposed a liquid-deuterium cryotarget and solid targets (carbon, iron, or lead) to the beam.  The data used in this work were taken in 2004, during the ``EG2'' run period. CLAS was a six-fold symmetric spectrometer that included drift chambers, time-of-flight (TOF) scintillation counters, gas Cherenkov counters, and an electromagnetic calorimeter.  Its polar angular acceptance ranged from 8$^\circ$ to 140$^\circ$ for the drift chambers and TOF, and from 8$^\circ$ to 45$^\circ$ for the Cherenkov counters and calorimeter. The drift chambers functioned in a toroidal magnetic field up to 2~Tm, where the magnet’s polarity made negative particles bend toward the beam pipe. The resulting resolutions for charged particles were $\sigma_p/p$=0.5\% for the momentum, 2~mrad for the polar angle, and 4~mrad for the azimuthal angle.

Electrons were identified by matching negatively charged tracks from the drift chambers with hits in the  TOF, the Cherenkov counter, and the calorimeter. Background suppression of $\pi^{-}$ to negligible levels was achieved by using the Cherenkov counter, the calorimeter, and the track momentum. Additionally, a fiducial selection based on drift chamber and calorimeter coordinate measurements was applied to the angle and momentum of the electrons, minimizing acceptance variations and energy leakage. Charged pions were identified by matching tracks from the drift chambers to hits in the TOF system. 

Fiducial selection criteria based on the pion laboratory-frame momentum, $p$, and polar angle, $\theta$, were applied: for $\pi^+$ we required $10^{\circ}<\theta<120^{\circ}$ and $p>200$~MeV.  
 For $\pi^-$, we required $p > 700$~MeV for $25^{\circ}<\theta<30^\circ$, 
$p > 500$~MeV for $30^{\circ}<\theta<40^{\circ}$, and   
$p > 350$~MeV for~$\theta>40^{\circ}$. The stricter selection criteria for $\pi^-$ reflect the decreased acceptance due to the toroidal field configuration that was used, which bent negatively charged particles towards the beam line. This approach, more restrictive than our prior double-ratio analysis~\cite{CLAS:2022asf}, aims to define a smoothly varying acceptance to better manage systematic uncertainties of the estimated correlation function.

The trigger system selected events with at least one electron candidate with $p>500$~MeV. Event selection was further refined offline, focusing on events with $Q^{2} > 1$~GeV$^{2}$, $W >2$~GeV, and $y < 0.85$. Here, $Q^{2}$ represents the square of the transferred four-momentum; $W$ is the invariant mass of the photon-nucleon system; and $y = \nu/E$, where $\nu$ is the transferred energy in the target rest frame and $E$ is the beam energy.

We selected events with a positively charged pion with an energy fraction $z\equiv E_h/\nu > 0.5$; we refer to this as the ``leading'' pion\footnote{The $z>0.5$ requirement  ensures that the leading pion is uniquely defined.}. We then measured all negatively charged pions with $z > 0.05$ in events containing a leading pion, which we term ``subleading'' pions.

To further reduce contamination from protons misidentified as $\pi^{+}$s, we required that the sum of the magnitudes of the hadron momenta be less than the energy transfer\footnote{The justification for this, as given in Ref.~\cite{CLAS:2022asf}, is that the sum of the energies of the hadrons produced in a reaction cannot exceed $\nu$. For knocked-out nucleons, only the kinetic energy counts towards this sum, since their rest energy is removed from the residual nuclear system.  This allows the knocked-out protons to have larger momenta than the pions for a given $\nu$.} (\textit{i.e.}, $|p_{1}| + |p_{2}| < \nu$). We also required  $p^T> 250$~MeV for each pion to mitigate the impact of acceptance effects with respect to $\Delta\phi$.

Particles arising from scattering off either the deuterium or nuclear targets were selected using the longitudinal vertex position. The resulting vertex resolution ensured negligible ambiguity in target assignment~\cite{CLAS:2021jhm, CLAS:2022asf}.

In all, there were 449k, 206k, 257k, and 877k events with a leading $\pi^+$ in the final event selection for the D, C, Fe, and Pb targets, respectively.  The number of $\pi^+\pi^-$ pairs for these targets were, respectively, 27.0k, 10.9k, 11.2k and 3.74k.

To obtain the correlation functions, we first divided the pion-pair sample into 8 bins of equal width in $|\Delta\phi|$, ranging from 0 to $\pi$,\footnote{We used the absolute value, $|\Delta\phi|$, instead of $\Delta\phi$, because we expect the values of the correlation function to be the same for both positive and negative values of $\Delta\phi$.}, and counted the number of pairs in each bin, $N^i_{e'\pi^+\pi^-}$, where $i$ is the bin number, as well as the total number of events with an electron and a leading pion $N_{e'\pi^+}$.  

Prior to corrections and normalization, the correlation function for the $i^{\rm th}$ bin is given by
\begin{equation}
    C^{i}_{\rm uncorr}=\frac{N^i_{e'\pi^+\pi^-}}{N_{e'\pi^+}},
\end{equation}  
where $N^i_{e'\pi^+\pi^-}$ is the number of pairs in the $i^{\rm th}$ bin and $N_{e'\pi^+}$ is the number of leading pions.  

The first correction we applied was for scattering off the 15~$\mu$m-thick aluminum cryotarget walls, which accounts for 2.5\% of the deuterium sample. This percentage was determined using data from an empty target and methodologies from Refs.~\cite{CLAS:2021jhm,CLAS:2022asf}.  We then corrected the deuterium correlation function by estimating the aluminum correlation function through interpolation from the measured carbon and iron correlation functions\footnote{In principle, the correlation function for aluminum could have been determined from the empty-target run; however, the statistical precision would be very poor in that case.}.  The corrected value of the deuterium correlation function was then
\begin{equation}
    C^i_{D,\rm cryo-corr} = C^i_{D,\rm uncorr}-0.025 (C^i_{\rm C,\rm uncorr}+C^i_{\rm Fe,\rm uncorr})/2,
\end{equation}
where the $D$, C, and Fe in the subscript refers to the target type.  
In the above procedure, we assumed that the correlation function for aluminum is halfway between carbon and iron.  The correlation functions for the nuclear targets are unchanged at this stage:
\begin{equation}
    C^i_{A,\rm cryo-corr} = C^i_{A,\rm uncorr},
\end{equation}
where the subscript $A$ indicates any nuclear target.

To determine correction factors arising from pair-acceptance, we used a data-driven ``mixed events'' method. This approach involved pairing the electron and leading pion from a single event with the subleading pion from various events within the same sample. By construction, this test sample exhibited no inherent physical correlation between the two pions, except for those correlations induced by detector acceptance effects. We determined the number of mixed events in each bin $N^{i, \rm mixed}_{e'\pi^+\pi^-}$ in $|\Delta\phi|$, and defined our correction factor for each bin to be
\begin{equation}
    M^i=\frac{N^{i, \rm mixed}_{e'\pi^+\pi^-}}{\sum\limits_{j=1}^{n_{\rm bins}} N^{j, \rm mixed}_{e'\pi^+\pi^-}/n_{\rm bins}},
\end{equation}
where $n_{\rm bins}$ is the number of bins (8) in $|\Delta\phi|$.  The correction factor $M^i$ for each bin was calculated using the deuterium sample and was uniformly applied across all targets\footnote{We used deuterium to determine the correction factor for all four targets because the efficiency of detecting a pair of particles should not depend on the type of target the particles originated from, and the deuterium dataset had the smallest statistical uncertainty of the four targets’ datasets.}, using
\begin{equation}
    C^i_{\rm acc-corr} = C^i_{\rm cryo-corr}/M^i.
\end{equation}

The final step in determining the correlation functions was normalization with respect to the deuterium measurement.  The final, normalized correlation functions were then
\begin{equation}
    C^i = \frac{C^{i}_{\rm acc-corr}}{\frac{2\pi}{n_{\rm bins}}\sum\limits^{n_{\rm bins}}_{j=1} C^{j}_{D,\rm acc-corr}}.\label{eq:norm}
\end{equation}

We then calculated the ratios for each bin, $i$, as
\begin{equation}
    R^i_A=C^i_A/C^i_D.
\end{equation}
where the subscripts $A$ and $D$ represent nuclear and deuterium measurements, respectively.  We determined the RMS
widths, $\sigma$, as
\begin{equation}
    \sigma=\sqrt{\frac{\sum\limits_{i=1}^{n_{\rm bins}} C^i(\Delta\phi^i-\pi)^2}{\sum\limits_{i=1}^{n_{\rm bins}} C^i}},
    \label{eq:sigmaDef}
\end{equation}
where $\Delta\phi^i$ is the central value of $\Delta\phi$ in the $i^{\rm th}$ bin, and used Eq.~\ref{eq:broadening} to determine the broadening, $b$.

\section{Systematic Uncertainties} \label{sec:systematics}
The systematic uncertainties in this measurement arise from procedures such as particle identification, target identification, event selection, as well as from pair-acceptance effects. To estimate these uncertainties, we conducted a series of data-driven and simulation studies. Some of these involved using the \textsc{Pythia 6.319} Monte-Carlo (MC) event generator~\cite{Sjostrand:2003wg} and the \textsc{GSIM} package~\cite{GSIM}, based on \textsc{Geant3}~\cite{Brun:1994aa}, to accurately simulate the CLAS detector response, including the dual-target configuration~\cite{Hakobyan:2008zz}.

\subsection{Particle Misidentification}
To estimate the effect of misidentification of charged-pion candidates, simulations with a deuterium target were conducted.    
We define the misidentification rate as
\begin{equation}
    f_{\rm misid}(\rm bin)=N_{\rm misid}(\rm bin)/N_{\rm tot}(\rm bin),
\end{equation} 
where $N_{\rm tot}$ is the total number of di-pion events that pass our event-selection criteria in a given kinematic bin, and $N_{\rm misid}$ is the number of those events in which one of the particles identified by the event reconstruction as a charged pion is actually something else (kaon, proton, \textit{etc.}).  We note that the timing resolution of the TOF in this simulation matches that of the data, which is necessary for an accurate estimation of the misidentification rates.
No misidentification of sub-leading $\pi^-$ was observed, but some kaons and protons were wrongly identified as leading $\pi^+$. Contamination rates were generally low (median 0.8\%), but higher (up to 14.4\%) in bins with fewer true $\pi^+\pi^-$ pairs. 

The nuclear targets are expected to have different ratios of the cross sections for producing background particles (kaons and protons) to those of our signal particles (charged pions) than for deuterium. We used the \textsc{GiBUU} model~\cite{Buss:2011mx} to estimate the difference in these cross-section ratios, allowing us to adjust the contamination rates for nuclear targets. We did this by scaling the estimated contamination rate for kaons (protons) in deuterium by the double ratio of kaons (protons) per pion in the nuclear targets to that of deuterium. The estimated numbers of protons and kaons are up to 2 to 3 times higher for lead than for deuterium.

Estimated contamination rates, $f_{\rm misid}$, were propagated as relative uncertainties in the correlation functions. For derived observables, uncertainties were accounted for by recalculating these observables with the estimated contamination subtracted from the correlation function (\textit{i.e.}, substituting $C(\Delta\phi)$ with $C(\Delta\phi)(1-f_{\rm misid}(\Delta\phi))$) and then taking the difference in these observables from the nominal values. Explicitly, 
\begin{equation}
    \Delta R(\Delta\phi) = \frac{C_A(\Delta\phi)(1-f_{{\rm misid},A}(\Delta\phi))}{C_D(\Delta\phi)(1-f_{{\rm misid},D}(\Delta\phi))}-\frac{C_A(\Delta\phi)}{C_D(\Delta\phi)},
\end{equation}
where the subscript $A$ or $D$ represents the quantity being evaluated for the nuclear or deuterium target, 
\begin{multline}
    \Delta \sigma = \sqrt{\frac{\sum_i  C^i(1-f_{i,\rm misid})(\Delta\phi_i-\pi)^2}{\sum_i C^i(1-f_{i, {\rm misid})}}}\\-\sqrt{\frac{\sum_i  C^i(\Delta\phi_i-\pi)^2}{\sum_i C^i}},
\end{multline}
and 
 \begin{multline}
    \Delta b = \sqrt{\begin{aligned}\frac{\sum_i  C_A^i(1-f^A_{i,\rm misid})(\Delta\phi^i-\pi)^2}{\sum_i C^A_i(1-f^A_{i, {\rm misid}})}\,\,\,\,\,\,\\
    -\frac{\sum_i  C^D_i(1-f^D_{i,\rm misid})(\Delta\phi_i-\pi)^2}{\sum_i C^D_i(1-f^D_{i, {\rm misid}})}\end{aligned}}\\
    -\sqrt{\frac{\sum_i  C^A_i(\Delta\phi_i-\pi)^2}{\sum_i C^A_i}
    -\frac{\sum_i  C^D_i(\Delta\phi_i-\pi)^2}{\sum_i C^D_i}}.
 \end{multline}

The values of these uncertainties varied by target and kinematic bin, with the largest uncertainty being 21.5\% for $R$, 3.0\% for $\sigma$, and 8.9\% for $b$. 

\subsection{Pair Acceptance}
Because of the way the correlation functions are normalized (see Eq.~\ref{eq:norm}), they are insensitive to variations in the overall efficiency (that is, the combined effects of geometric acceptance and detector efficiency) that would affect all kinematic bins by the same factor, and are instead sensitive to relative bin-by-bin variations of the efficiency.
Systematic uncertainties, stemming from pair acceptance affecting $C(\Delta\phi)$, were estimated using two methods. 
In the first method, we used MC simulations, and we estimated the overall efficiency of our detector  on a bin-by-bin basis as the ratio of the number of simulated events in each $\Delta\phi$ bin in which all three particles were reconstructed and pass event-selection cuts to the total number of generated events in that bin.  The systematic uncertainty of the correlation function from this method was then the ratio of the standard deviation of these efficiencies to the mean value, yielding 3.6\%.
The second method involved the mixed-event method that we used for correcting our data. The systematic uncertainty from this method was the standard deviation of the values of $M(\Delta\phi)$ over all $\Delta\phi$ bins, yielding 6.2\%.  Conservatively, we used the larger of these two estimates.  

The difference in the acceptance between different targets was evaluated to be $\approx$2\% in the context of an earlier analysis~\cite{CLAS:2022asf}, which is much smaller than our estimate of bin-by-bin variations in the acceptance.  Therefore, we used the same value for the uncertainty for each target.  
To further verify the pair-acceptance corrections, we reassessed our $M(\Delta\phi)$ estimates while requiring that events only be mixed together with events that have similar $Q^2$ and $x$ values. This additional requirement had a negligible impact on the results.

For $R$, there is some partial cancellation of the pair acceptance from target to target, and we estimated a residual 2\% uncertainty, as was determined in the context of an earlier analysis~\cite{CLAS:2022asf}.  
To propagate uncertainties to $\sigma$ and $b$, we employed a numerical pseudosample method with bin-by-bin variations and recalculated the observables. This approach determined the systematic uncertainty for $\sigma$ across each target and kinematic bin to be between 0.9--1.4\%. For $b$, this resulted in a standard deviation of 1.4\%, which we adopted as the uncertainty for all targets and kinematic bins.

\subsection{Event Selection}
To assess the impact of the event-selection cuts, we revisited our data with various cut values, as outlined in Ref.~\cite{CLAS:2022asf}, focusing on electron, pion, and vertex selection. Only alterations to the minimum $p^T$ cut significantly affected our results. The systematic uncertainty on $C$ was calculated as  
\begin{equation}
    \frac{\Delta C}{C} = \sqrt{\frac{1}{N_{\rm var}}\sum\limits_{\rm var}\left(\sum\limits_{\rm targets}\frac{C_{\rm nom}-C_{\rm var}}{C_{\rm nom}}\right)^2},
    \label{eq:uncCuts}
\end{equation}
where the subscript ``nom'' refers to the nominal cuts ($p^T$>250 MeV), ``var'' refers to the variants of the cuts ($p^T$>230 MeV for the looser variant, and $p^T$>270 MeV for the tighter variant), $N_{\rm var}$ is the number of cut variations (2), and $N_{\rm targets}$ is the number of targets (4).  We calculated the uncertainties of $R$, $\sigma$, and $b$ by replacing $C$ in Eq.~\ref{eq:uncCuts} with $R$, $\sigma$ or $b$, and setting $N_{\rm targets}$ to 3 for $R$ and $b$ (since we only evaluate those observables for the nuclear targets).  
The relative systematic uncertainties were up to 11.4\% for the correlation function and up to 21.5\% for the nuclear-to-deuterium ratio, with median values of 3.2\% and 1.4\%, respectively. For $\sigma$ ($b$), the relative systematic uncertainty from event-selection cuts was in the range 0.9\% to 2.5\% (0.3\% to 3.8\%).  These uncertainties were calculated for each kinematic bin, with identical values applied across all targets. 
    
\subsection{Finite Bin Width}
The use of binned data is a source of a systematic error for $\sigma$, due to variations within each bin. We assessed this by analyzing unbinned toy Gaussian distributions with varying standard deviations, comparing RMS values from the binned distributions, that is,
\begin{equation}
    \sigma_{\rm binned}=\sqrt{\frac{1}{N_{\rm tot}}\sum\limits_{j\in\rm bins} N_j (\Delta\phi^{\rm bin-center}_j-\pi)^2},
\end{equation}
to those from the unbinned distributions, \textit{i.e.,}
\begin{equation}
    \sigma_{\rm unbinned}=\sqrt{\frac{1}{N_{\rm tot}}\sum\limits_{i\in events} (\Delta\phi_i-\pi)^2}.
\end{equation}
Here, $N_{\rm tot}$ is the total number of events in the toy distribution, $N_j$ is the number of events in the $j^{\rm th}$ bin, $\Delta\phi^{\rm bin-center}_j$ is the central value of $\Delta\phi$ in that bin.  In the unbinned case, the sum is over all events, and $\Delta\phi_i$ is the value of $\Delta\phi$ in the $i^{\rm th}$ event.  We then determined the average discrepancy between the binned and unbinned values, and found that this average discrepancy depended on the width of the distribution (with a larger discrepancy for distributions with smaller widths relative to the width of a bin). We then ran a fit of these average discrepancies as a function of the original distribution’s width.  Since we did not make any assumption about the shape of the distributions in our data, we took as our uncertainty the estimated value of this offset determined for Gaussian functions.  We found that for values of $\sigma$ comparable to those measured in our analysis, the resulting value of the uncertainty is up to 1.4\%.  

For the broadenings, we did a similar analysis with pairs of toy distributions with different widths (with the narrower distribution acting as a proxy for the deuterium correlation function and the wider acting as a proxy for the nuclear correlation function). The ``broadenings'' of these pairs were the quadrature difference between the widths in the pair of distributions. For each pair of distributions, we also calculated the broadenings using the binned RMS widths of the distributions. The average discrepancy between the broadening obtained with the binned and unbinned methods was small, about 3~mrad.  Again, since we do not make an assumption about the shapes of the distributions in our data, we assigned this value as the systematic uncertainty on the broadening for every target and kinematic bin. For the values of $b$ we determined in this work, this systematic uncertainty is about 0.4\% to 0.7\% of the measured values of $b$.

\subsection{Cryotarget Walls}
To account for background events in our event sample stemming from scattering off the aluminum cryotarget walls, we used the method described in Sec.~\ref{sec:setup}. We assumed that the correlation function for aluminum is halfway between those of carbon and iron, however this introduces some small systematic uncertainty,

\begin{equation}
    \Delta C_D = f_{\rm Al} \sqrt{(\Delta C_{\rm C})_{\rm stat}^2+(\Delta C_{\rm Fe})_{\rm stat}^2+(C_{\rm C}-C_{\rm Fe})^2}/2,
    \label{eq:deltaC_D_EC}
\end{equation}
where the first two terms arise from the statistical uncertainty for the carbon and iron correlation functions, respectively, and the last term accounts for the uncertainty on the interpolation itself.  This yields a 0.1\% to 2.6\% uncertainty on the correlation function for deuterium, depending on the bin.  Since this correction to the deuterium correlation affects the normalization factor being used, it also indirectly affects the nuclear-target data as well, resulting in a 0.6\% to 0.8\% scale uncertainty in the correlation functions.  The systematic uncertainty of $R$ includes both the systematic uncertainty from the shape of the deuterium correlation function (which is correlated between nuclei) and the normalization uncertainty from the nuclear targets.  The estimated systematic uncertainty was up to 0.1\% for $\sigma$ and up to 0.2\% for $b$. 

\subsection{Other Sources of Systematic Uncertainty}
We found through simulations that the detector resolution's effect on bin migration is negligible, as the resolutions in our binning variables are significantly smaller than the smallest bins. 

Additionally, luminosity and trigger efficiency do not contribute to the systematic uncertainty of the correlation function as these affect the numerator and denominator of Eq.~\ref{eq:defC} equally. Also, time-dependent effects were found to be negligible, evidenced by consistent results for the correlation functions (within statistical uncertainties) across three subdivisions of the deuterium target's total run period. 

We also compared the values of the correlation function for every bin at $\Delta\phi$ with the one at $2\pi-\Delta\phi$, and found the difference between the two to be negligible (as expected from symmetry considerations).  

The effect of misidentifying other particles\footnote{such as the $e^-$ from background lepton pairs or $\pi^-$ identified as electrons} as scattered electrons was investigated in the context of previous measurements from the same dataset (such as Refs.~\cite{CLAS:2021jhm,CLAS:2022asf}), and found to be negligible.  Likewise, radiative effects were also found to be negligible in the context of those measurements.  

\subsection{Summary of Systematic Uncertainties}
We summarize the statistical and systematic uncertainties from the various sources described above in Table~\ref{tab:syst_C} for the correlation functions and their nuclear-to-deuterium ratios, and in Table~\ref{tab:syst_sigma} for the widths and broadenings. Unless specified, each systematic uncertainty is estimated to vary point-to-point between bins. Notably, some bins have systematic uncertainties significantly larger than others. The median uncertainty per bin, which is 7.4\% (7.3\%) for $C(\Delta\phi)$ for $A$ ($D$), 2.7\% for $R$, 2.5\%  (2.4\%) for $\sigma$ for $A$ ($D$), and 3.6\% for $b$.  
\setlength{\tabcolsep}{12pt}
\begin{table*}[]
    \centering
    \begin{tabular}{c|c c c c c}
        Source & $\Delta C/C$ ($D$) & $\Delta C/C$ ($A$)  & Type & $\Delta R/R$ & Corr. $A$ vs $D$?\\
        \hline
        Statistics & 1.1$-$38.8\% &  1.8$-$43.8\% & p2p &  2.2$-$52.7\% & N \\
\hline
Particle misid. & 0$-$14.3\% &  0$-$32.7\%& p2p &  0$-$21.5\%  & Y \\
Pair acceptance & 6.2\% &  6.2\% & p2p &  2.0\%  & Y\\
Event selection & 0.2$-$11.4\% &  0.2$-$11.4\%& p2p &  0.1$-$21.5\%  & Y \\
Cryotarget walls & 0.1$-$2.6\%  & -- & p2p & 0.1$-$2.6\% &  --\\
Cryotarget walls (renorm) & -- &  0.6$-$0.8\%  & norm & 0.6$-$0.8\% & -- \\
Bin migration & negligible & negligible & -- & negligible & --\\
Luminosity & negligible & negligible & -- & negligible & -- \\
Time-dependent effects & negligible  & negligible  & -- & negligible & --\\
Trigger efficiency & negligible & negligible  & -- & negligible & --\\
\hline
Syst. subtotal & 6.2$-$16.5\% &  6.3$-$33.7\%  & -- &  2.1$-$22.2\% & Y\\
\hline
Total & 6.4$-$40.1\% &  6.7$-$47.9\% & -- &  3.1$-$53.1\% & Y \\

    \end{tabular}
    \caption{Summary of the statistical and systematic uncertainties of the correlation functions of deuterium ($D$) and nuclear targets ($A$). These include uncertainties for the ratio $R=C_A/C_D$ and the correlation of systematic uncertainties between $D$ and $A$ targets. Uncertainties are categorized as ``point-to-point'' (p2p), affecting each bin differently (with possible bin correlations), and ``normalization'' (norm), impacting all bins uniformly. The column marked ``corr.~$A$ vs $D$'' indicates whether or not the systematic uncertainties for the nuclear measurement and that of the deuterium measurement are correlated.  
    }
    \label{tab:syst_C}
\end{table*}

\begin{table*}[]
    \centering
    \begin{tabular}{c|c c c c}
        Source & $\Delta \sigma/\sigma$ ($D$) & $\Delta \sigma/\sigma$ ($A$) & Corr.~$D$ vs $A$? & $\Delta b/b$ \\
        \hline
        Statistics & 0.4$-$2.5\% &  0.6$-$4.9\% & N &  2.7$-$14.1\% \\
\hline
Particle misid. & 0.1$-$0.8\% &  0$-$3.0\% & Y &  0.3$-$8.9\% \\
Pair acceptance & 1.0$-$1.4\% &  0.9$-$1.2\% & Y &  1.4\% \\
Event selection & 0.9$-$2.5\% &  0.9$-$2.5\% & Y &  0.3$-$3.8\% \\
Finite bin width & 0.1$-$1.4\% &  0$-$0.7\% & Y &  0.4$-$0.7\% \\
Cryotarget walls & 0$-$0.1\% &  -- & N &  0$-$0.2\% \\
Bin migration & negligible & negligible & -- & negligible \\
Luminosity & negligible & negligible & -- & negligible \\
Time-dependent effects & negligible & negligible & -- & negligible \\
Trigger efficiency & negligible & negligible & -- & negligible \\
\hline
Syst. subtotal & 1.9$-$3.2\% &  2.0$-$3.7\% & Y &  2.4$-$9.2\% \\
\hline
Total & 2.0$-$3.8\% &  2.2$-$5.4\% & Y &  4.5$-$14.8\% \\

    \end{tabular}
    \caption{Summary of statistical and systematic uncertainties for widths and broadenings, detailed separately for deuterium ($D$) and nuclear targets ($A$). The third column indicates whether the uncertainties for $\sigma$ in the $D$ and $A$ targets are correlated with one another.}
    \label{tab:syst_sigma}
\end{table*}
\section{Results}
\label{sec:Results}
\begin{figure}[h!]
    \centering
    \includegraphics[width=0.95\columnwidth]{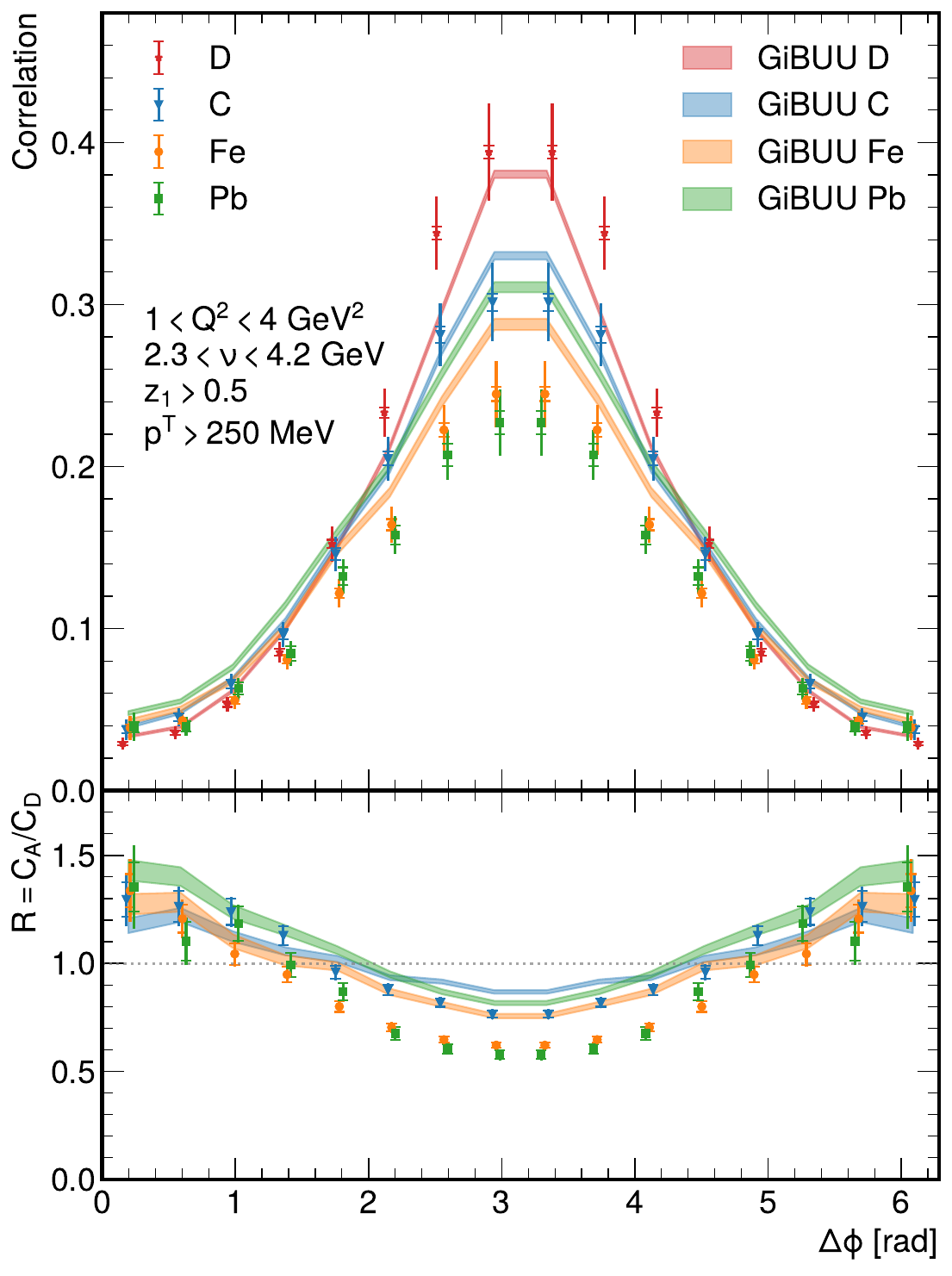}
    \caption{Top: Correlation functions for all four targets. Bottom: Ratio of nuclear target correlations to deuterium. Data points are slightly shifted horizontally for clarity. Data are compared to calculations from the \textsc{GiBUU} model, shown as curves with widths representing the statistical precision of the \textsc{GiBUU} simulation. The inner error bars indicate statistical uncertainties, and the vertical extent of the error bars reflects the total uncertainty, combining systematic and statistical uncertainties (quadrature addition).}
    \label{fig:correlations}
\end{figure}

The top panel of Fig.~\ref{fig:correlations} shows $C(\Delta\phi)$\footnote{We note that we measured the correlation functions and ratios as a function of $|\Delta\phi|$ in the range of 0 to $\pi$.  For illustration purposes we show this as a function of $\Delta\phi$, ranging from 0 to $2\pi$ by showing the values for $0<\Delta\phi<\pi$, and shifting the points at $-\pi<\Delta\phi<0$ by $2\pi$, so that they are from $\pi<\Delta\phi<2\pi$ instead.}, integrated over di-pion kinematics with $p^{T}>250$ MeV for both pions. The functions peak around $\Delta\phi=\pi$, a consequence of momentum conservation, and have less strength and are broader in heavy nuclei compared to deuterium data. In the tails, the correlation functions for nuclear targets show a significant increase compared to deuterium, with a relative rise of up to about 40\%, as detailed in the bottom panel of Fig.~\ref{fig:correlations}, which shows the nuclear-to-deuterium ratio.

We compare our results with calculations from the \textsc{GiBUU}~model~\cite{Buss:2011mx}, which accounts for final-state interactions, absorption, and production mechanisms, including both elastic and inelastic channels.  The version of the \textsc{GiBUU}~model used was the 2021 release.  We modified the default values of several parameters related  to parton distributions and fragmentation in free nucleons, based on a tuning study~\cite{ElAlaoui_PC} that used an independent proton-target dataset from Ref.~\cite{CLAS:2022sqt}.  This tuning does not modify the nuclear effects.

The model shows close alignment with the deuterium data (top panel of Fig.~1), but notable numerical discrepancies emerge for nuclear targets, particularly in the peak heights of heavier nuclei. The peak heights in the data decrease with nuclear size, whereas the peak height produced by the model is higher for Pb than for Fe. Pb's higher neutron count, enhancing $\pi^-$ production, suggests the \textsc{GiBUU}~model may be overcompensating for $\pi^-$ production off the neutrons.

\begin{figure}[h!]
    \centering
    \includegraphics[width=0.95\columnwidth]{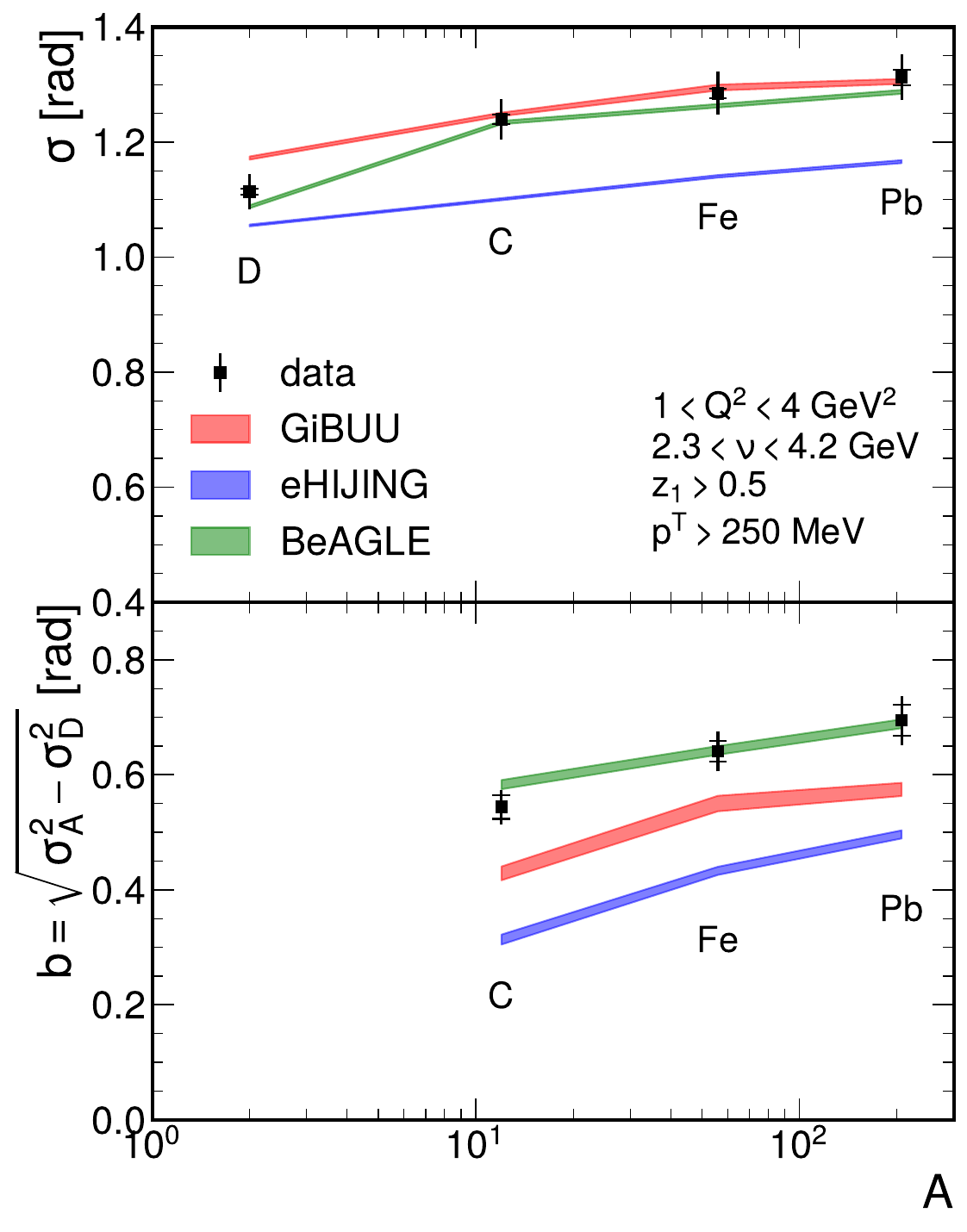}
    \caption{Top: Correlation function widths for all four targets. Bottom: Relative broadenings of nuclear targets compared to deuterium. The inner error bars represent statistical uncertainties, while the vertical extent of the error bars shows the combined systematic and statistical uncertainties (quadrature addition). These are compared with calculations from the \textsc{GiBUU}~(red), \eHIJING~(blue), and \BeAGLE~(green) models, where the widths of the curves indicate the statistical precision of each model's simulation.}
    \label{fig:widths_1d}
\end{figure}

We show the widths of the correlations, $\sigma$, in the top panel of Fig.~\ref{fig:widths_1d}.  In the bottom panel of Fig.~\ref{fig:widths_1d}, we show the broadening of the correlation functions for nuclei relative to that of deuterium (see Eq.~\ref{eq:broadening}).  The broadenings are 0.54$\pm$0.02$\pm$0.02~rad for C, 0.64$\pm$0.02$\pm$0.03~rad for Fe, and 0.70$\pm$0.03$\pm$0.03~rad for Pb, where the second and the third values denote the statistical and the systematic uncertainties, respectively.  The dependence of the widths and broadenings on the atomic mass number, $A$, appears to be at most logarithmic and not a power law.  

Naively, one would assume that the amount of angular broadening per unit length of travel through the nuclear medium would be fixed, and that the average path length for a particle escaping the nucleus would be proportional to the nuclear radius.  From these assumptions, it would follow that the azimuthal broadening would scale with the nuclear radius (which is proportional to $A^{1/3}$), however, this is not supported by the data.  

The \GiBUU~model accurately reproduces the widths for all three nuclear targets within the uncertainties, but it overestimates the width for deuterium by approximately 5\%. Due to this discrepancy, the calculated broadenings are around 20\% lower than the observed data. However, the model does reflect a trend of increasing broadening with larger nuclear sizes, similar to the trend observed in the data.

We also compared our results for the widths and broadenings with calculations from the \eHIJING~model~\cite{Ke:2023xeo} version 1.0 and the \BeAGLE~event generator~\cite{Chang:2022hkt} version 1.01.03\footnote{A comparison between  these model calculations for the correlation functions can be found in App.~\ref{app:more_models}.}. In \eHIJING, the hard-scattering process, along with initial- and final-state fragmentation, is simulated using \textsc{Pythia8}~\cite{bierlich2022comprehensive}. The model incorporates nuclear modifications to parton-distribution functions (PDFs), fragmentation, and parton showering.  It also handles multiple collisions between partons and the nucleus, with a cross section proportional to the transverse-momentum-dependent gluon distribution density. The \BeAGLE~event generator integrates modules from various sources. The primary interaction is managed by \textsc{Pythia6}~\cite{Sjostrand:2006za}. It incorporates nuclear PDFs from \textsc{LHAPDF5}~\cite{Whalley:2005nh}, intranuclear scattering via \textsc{DPMJet}~\cite{Roesler:2000he}, and the geometric density of nucleons through \textsc{PyQM}~\cite{Dupre:2011afa}. Nuclear remnant de-excitations and decays are processed using \textsc{FLUKA}~\cite{Ferrari:2005zk}.  In both the \eHIJING~and \BeAGLE~event generators, we used the same set of fine-tuned parameters based on the proton-target data described above as we had used for \textsc{GiBUU}.  

\begin{figure*}
    \centering
    \includegraphics[width=\textwidth]{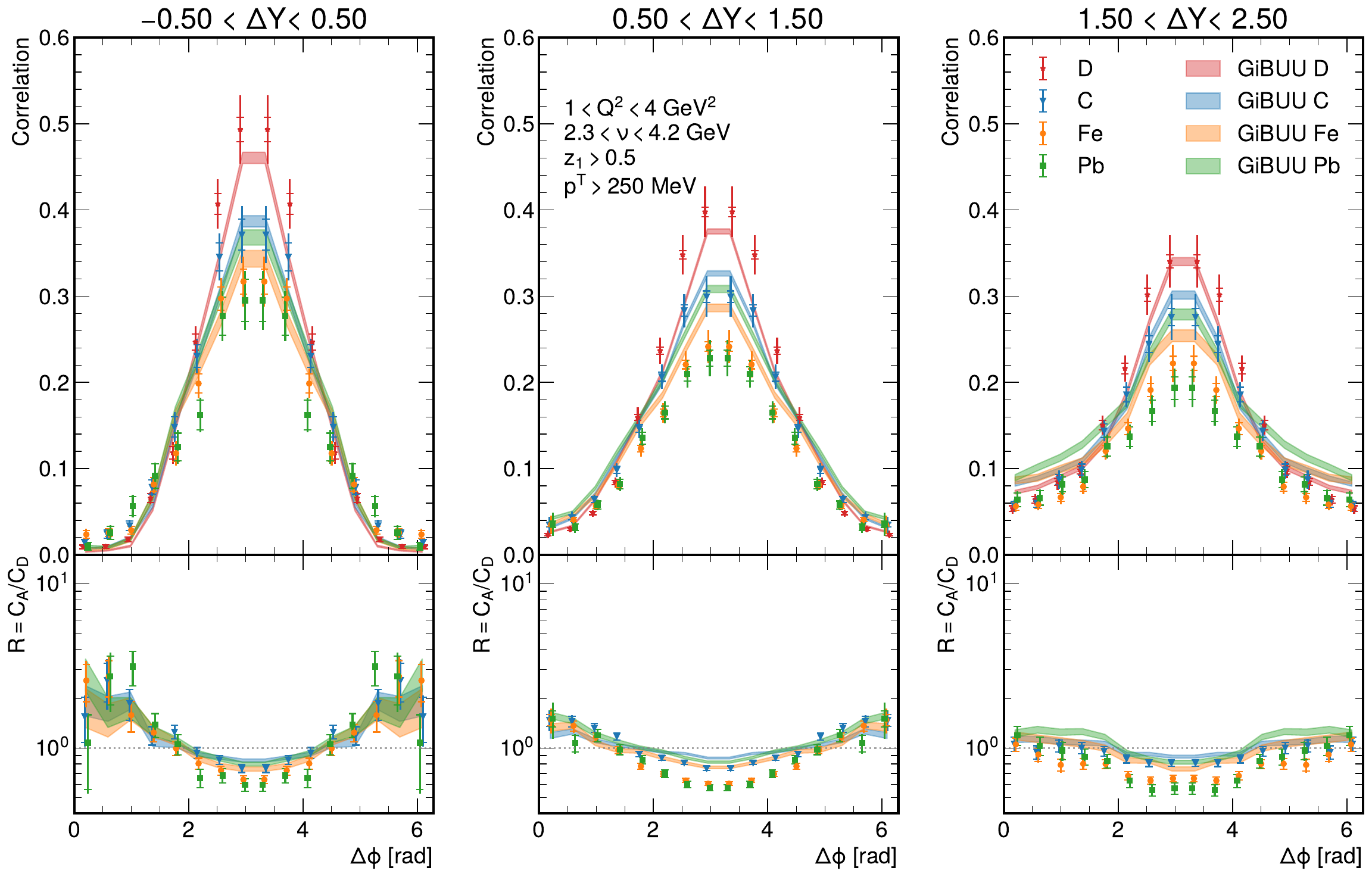}
    \caption{Top row:  Correlation functions in bins of rapidity difference, $\Delta Y\!$.  These are compared to the \textsc{GiBUU} calculations (curves).  The inner error bars represent the statistical uncertainties, while the vertical extent of the error bars represents the total uncertainty (the systematic and statistical uncertainties added in quadrature).  The width of the curves represents the statistical precision of the simulated sample.  Bottom: Ratios of the correlation functions for the nuclear targets to that of deuterium.}
    \label{fig:dY}
\end{figure*}

The widths calculated in the \eHIJING~model increase with larger nuclear size, however, the calculated widths are below those of the data, with a larger discrepancy for deuterium than in the nuclear data.  Further, the calculated broadenings are about 30\%--40\% lower than those in the data.  The values of the correlation widths and broadenings calculated with BeAGLE are consistent with our data within their uncertainties for each target.

\begin{figure*}
    \centering
    \includegraphics[width=0.85\textwidth]{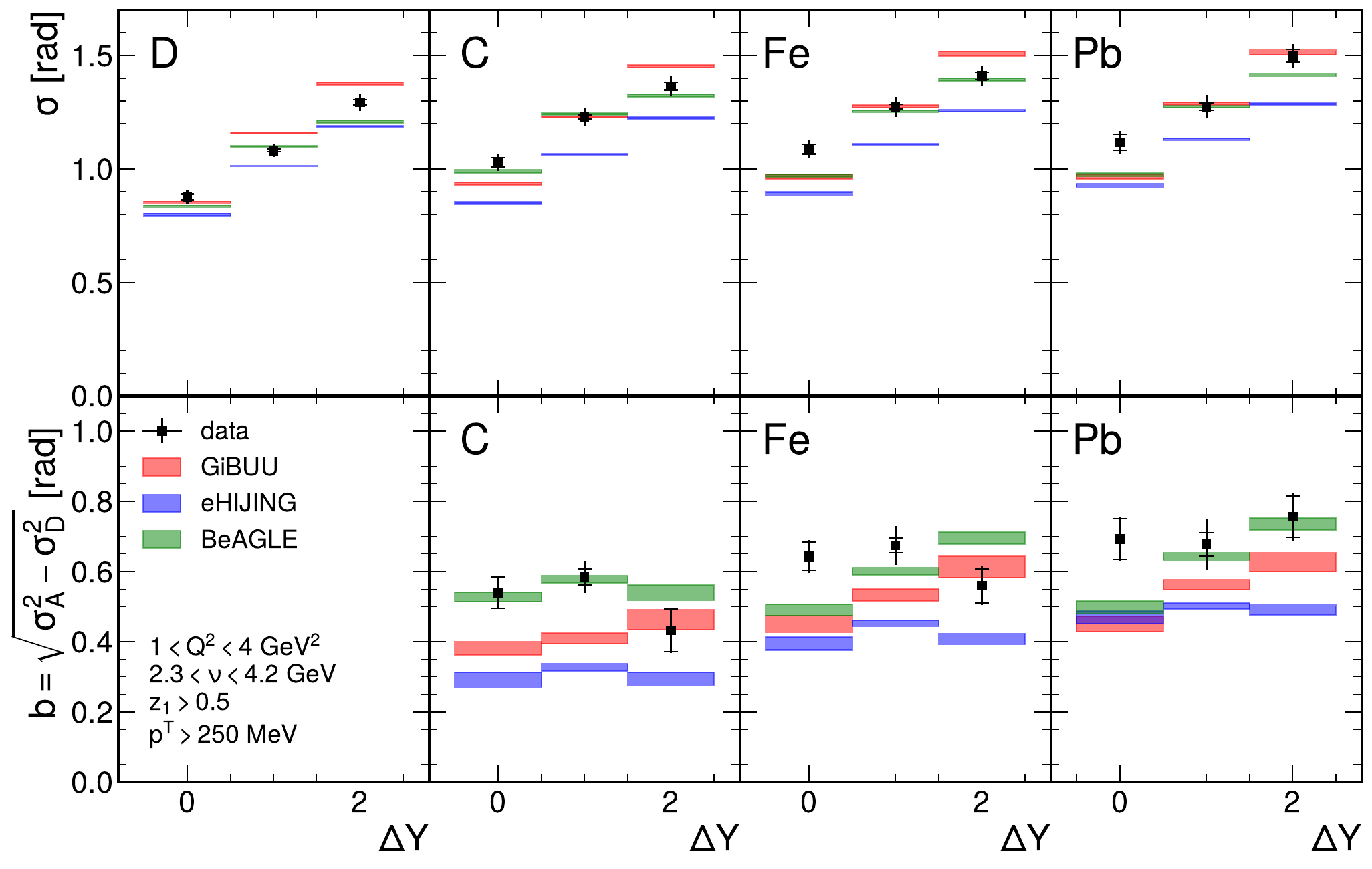}
    \caption{Top row:  RMS widths of the correlation functions in slices of $\Delta Y$ for each target, compared to the \GiBUU, \eHIJING,~and \BeAGLE~event generators.  Bottom row:  Same for the broadening.}
    \label{fig:dY_summary}
\end{figure*}

We show the correlation functions in slices of $\Delta Y$ in Fig.~\ref{fig:dY}.   The variables $\sigma$ and $b$, determined from these slices, are shown in Fig.~\ref{fig:dY_summary}.   We find that the correlation functions become wider with increasing $\Delta Y\!$.  Further, at low $\Delta Y\!$, the values of the nuclear-to-deuterium ratios in the tails of the correlation function ($\Delta\phi$ close to 0 or 2$\pi$) are nearly 3 times larger than unity.  At large $\Delta Y\!$, on the other hand, these ratios do not exceed 20\% higher than unity.

All three models show a increase in the RMS width with respect to increasing $\Delta Y\!$, similar to the data.  The \GiBUU~model's calculated values have a larger slope for $\sigma$ with respect to $\Delta Y$ than in the data, however, it is consistent with the data in the first bin for deuterium and the middle bins for each of the nuclear targets. 
Unlike the case of the integrated results, the \BeAGLE~correlation functions in bins of $\Delta Y$ exhibit some numerical discrepancies from the data.  The values for the middle bin (centered at $\Delta Y=1.5$) are consistent with the data within the uncertainties. The \eHIJING~model consistently underestimates the widths.

The experimental values of the broadening, $b$, show no apparent trend with respect to $\Delta Y\!$.  Both the \GiBUU~and \BeAGLE~event generators show an increase in $b$ with increasing $\Delta Y$ for Fe and Pb, which is inconsistent with the data.  The values of $b$ calculated by the \eHIJING~model are up to $\approx$50\%  smaller than in the data, however this model is the only one that does not yield a non-zero slope in $b$ with respect to $\Delta Y\!$.

\begin{figure*}
    \centering
    \includegraphics[width=\textwidth]{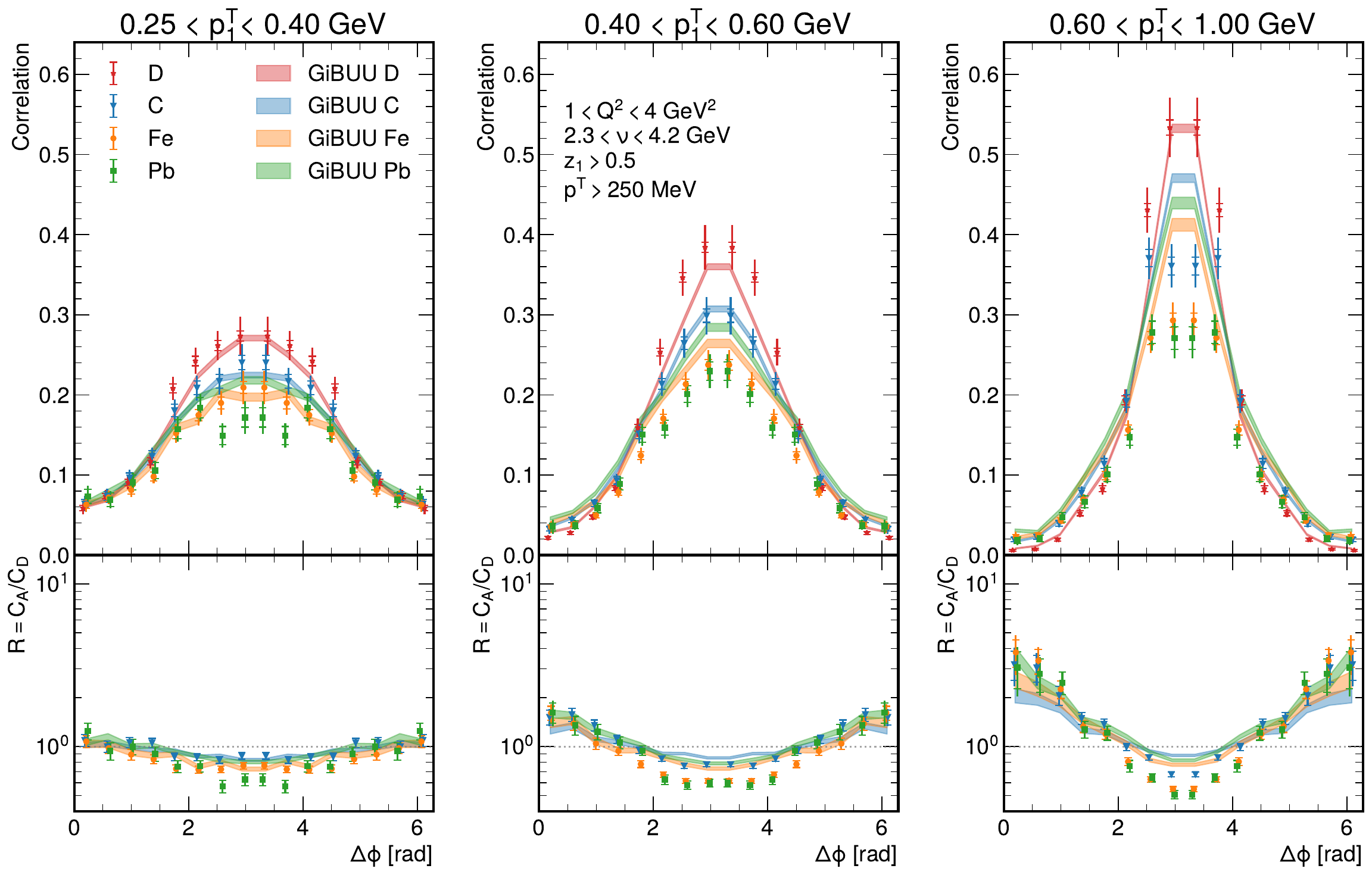}
    \caption{Top row:  Correlation functions in bins of the leading pion's transverse momentum, $p^T_1$.  These are compared to the \textsc{GiBUU} calculations (curves).  The inner error bars represent the statistical uncertainties, while the vertical extent of the error bars represents the total uncertainty (the systematic and statistical uncertainties added in quadrature).  The width of the curves represents the statistical precision of the simulated sample.  Bottom: Ratios of the correlation functions for the nuclear targets to that of deuterium.}
    \label{fig:pT1}
\end{figure*}

\begin{figure*}
    \centering
    \includegraphics[width=0.85\textwidth]{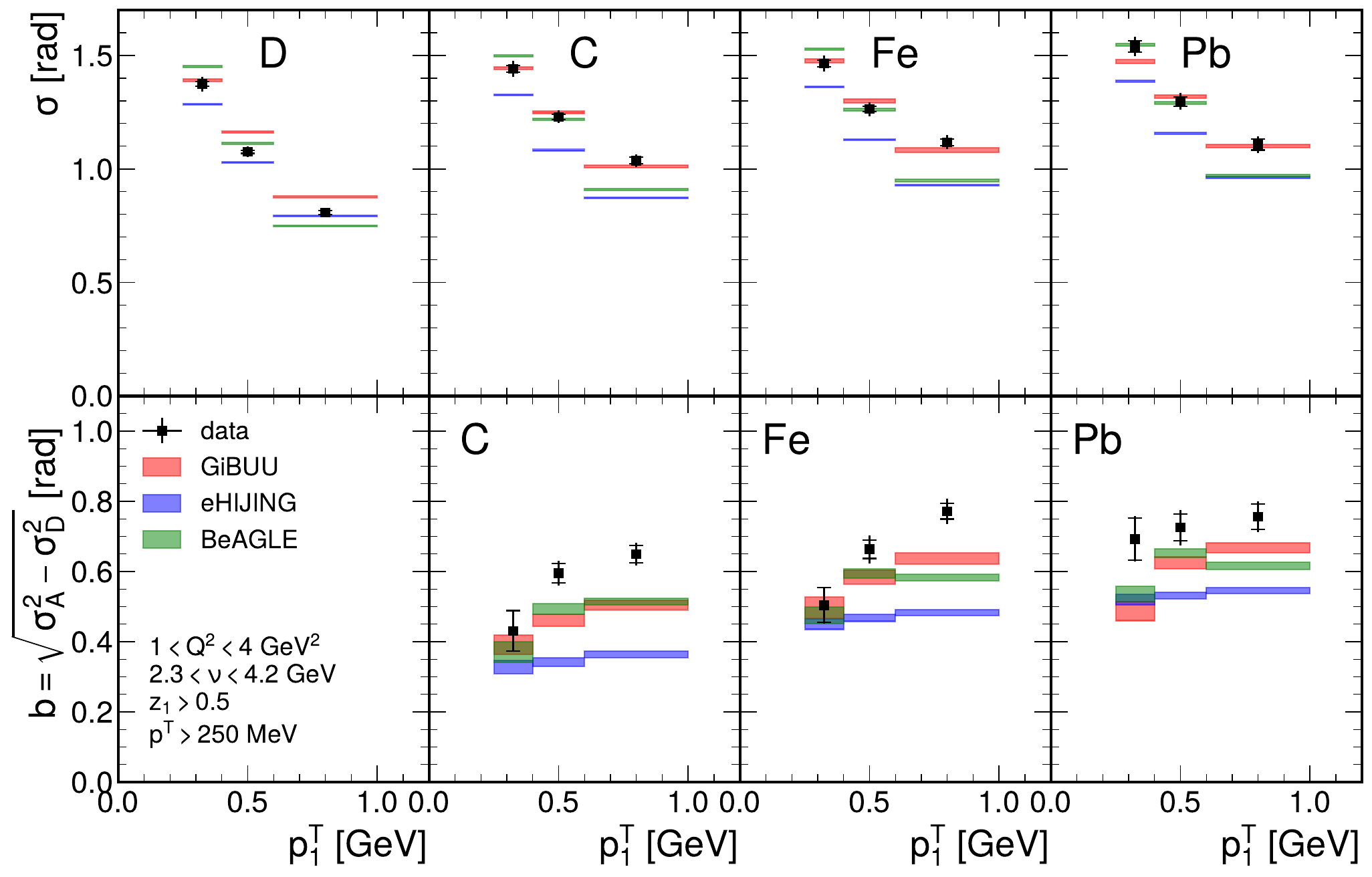}
    \caption{Top row:  RMS widths of the correlation functions in slices of $p^T_1$ for each target, compared to the \GiBUU, \eHIJING,~and \BeAGLE~event generators.  Bottom row:  Same for the broadening.}
    \label{fig:pT1_summary}
\end{figure*}

We likewise show the correlation functions in slices of $p^T_1$ in Fig.~\ref{fig:pT1}, and the variables $\sigma$ and $b$  for these slices in Fig.~\ref{fig:pT1_summary}.  At low $p^T_1$, the correlation functions are wide, whereas at high $p^T_1$, these functions are much narrower.  Further, at large $p^T_1$, the nuclear-to-deuterium ratios are up to $\approx$6.  

The calculations of all three models yield decreasing widths as $p_1^T$ increases.  The \GiBUU~model reproduces the width within one standard deviation ($1\sigma$) for all bins except for the first bin for Pb, and the second and third bins for D.  The \eHIJING~model shows a consistently smaller width than observed in the data.  The \BeAGLE~event generator shows a stronger dependence of the width on $p^T_1$ than observed in the data, with the calculation for the middle bin (centered at $p^T_1=0.5$~GeV) being close to or consistent with the data for that bin.  

The data show an increase in the broadening, $b$, with respect to $p^T_1$ for C and Fe.  However, for Pb, the dependence of $b$ on $p^T_1$ is smaller than the uncertainty on $b$. All three models also show an overall positive slope of $b$ with respect to $p^T_1$.  

\begin{figure*}
    \centering
    \includegraphics[width=\textwidth]{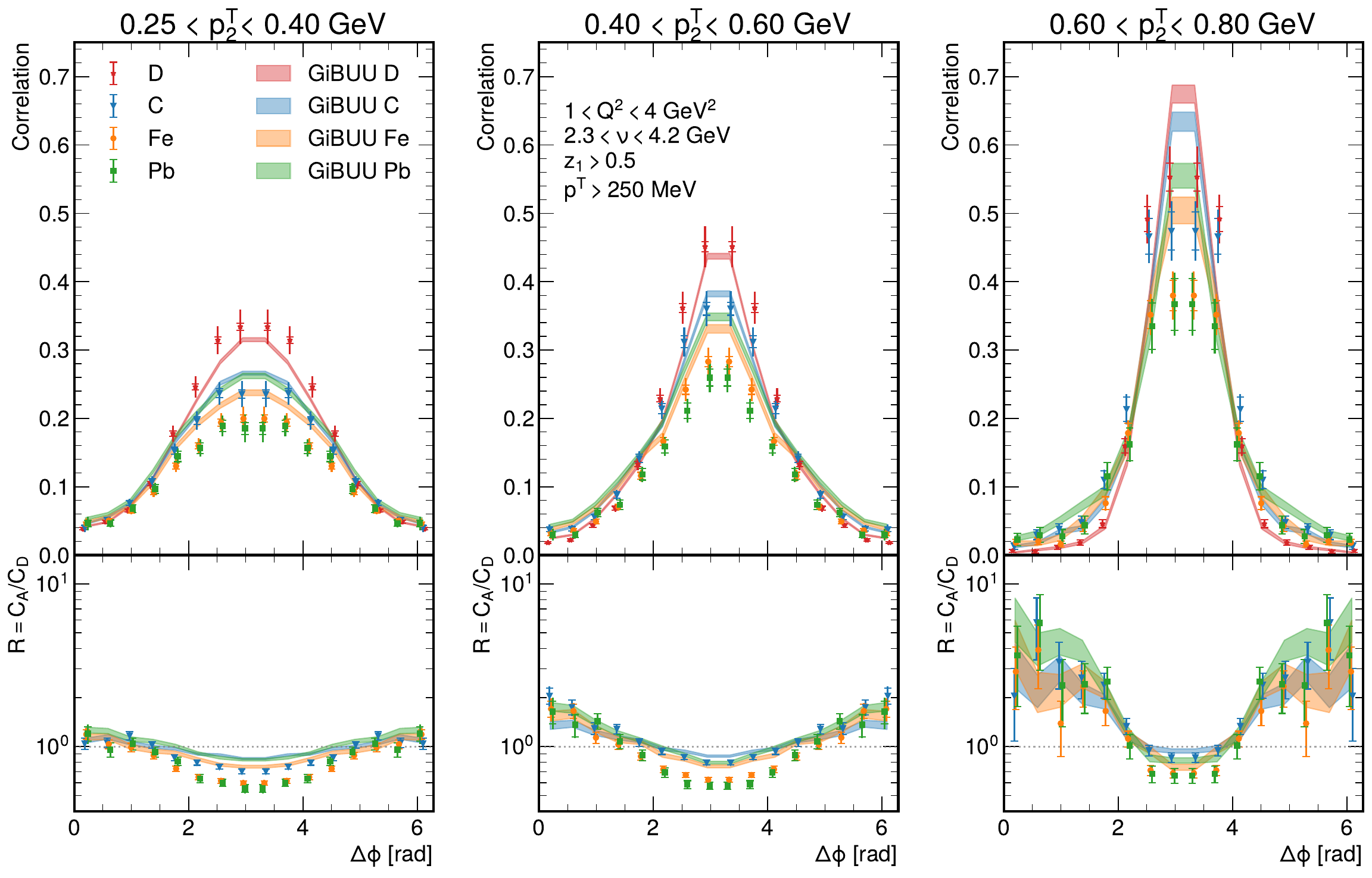}
    \caption{
    Top row:  Correlation functions in bins of the subleading pion's transverse momentum, $p^T_2$.  These are compared to the \textsc{GiBUU} calculations (curves).  The inner error bars represent the statistical uncertainties, while the vertical extent of the error bars represents the total uncertainty (the systematic and statistical uncertainties added in quadrature).   The width of the curves represents the statistical precision of the simulated sample.  Bottom: Ratios of the correlation functions for the nuclear targets to that of deuterium.}
    \label{fig:pT2}
\end{figure*}

\begin{figure*}
    \centering
    \includegraphics[width=0.85\textwidth]{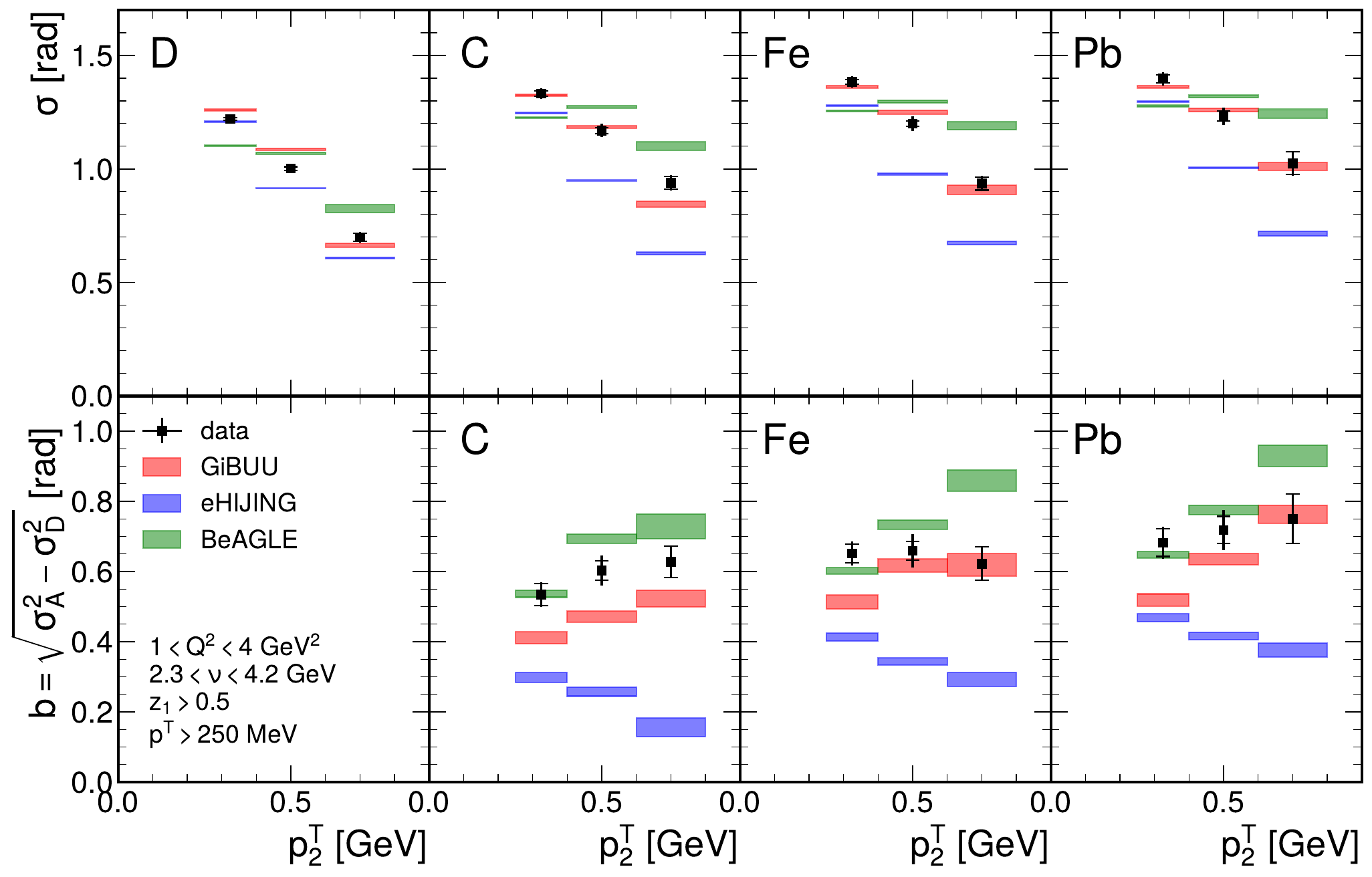}
    \caption{Top row:  RMS widths of the correlation functions in slices of $p^T_2$ for each target, compared to the \GiBUU, \eHIJING,~and \BeAGLE~event generators.  Bottom row:  Same for the broadening.}
    \label{fig:pT2_summary}
\end{figure*}

Finally, we show the correlation functions in slices of $p^T_2$ in Fig.~\ref{fig:pT2}, and the variables $\sigma$ and $b$ in these slices in Fig.~\ref{fig:pT2_summary}.  The correlation functions become narrower for larger $p^T_2$, and the values of the nuclear to deuterium ratios are around 6 at large $p^T_2$ and $\Delta\phi$ near 0 ($2\pi$).  

The \GiBUU~and \eHIJING~models show a decrease in the RMS widths with respect to $p^T_2$, qualitatively consistent with the data.  The \BeAGLE~event generator, on the other hand, shows a much weaker dependence on $p^T_2$, especially for Fe and Pb.  

Further, while the data show very little, if any, dependence of $b$ on $p^T_2$, \GiBUU~and \BeAGLE~show a positive slope for $b$ with respect to $p^T_2$, while \eHIJING~shows a negative slope.  

\section{Summary and Conclusions} \label{sec:conclusions}
We have presented the first measurement of di-pion azimuthal correlation functions in electron scattering off nuclei. This measurement reveals how two hadrons in a single scattering event are correlated, and its nuclear dependence brings qualitatively new data to probe how hadronization manifests in a heavy nucleus. 

The correlation functions peak at $\Delta\phi=\pi$ (opposite pairs) and are smaller near $\Delta\phi=0$ (close pairs), with RMS widths varying from 0.9--1.6 rad, depending on the target and kinematics. The correlations are wider at larger rapidity differences, but are narrower at higher transverse momentum (of either pion). We defined azimuthal broadening as the quadrature difference of these widths compared to deuterium, and found it to be a few hundred mrad. Our data show that the broadening increases weakly as the atomic mass number increases. This observable shows moderate kinematic dependence.

A comparison of our results with calculations from three models, namely \textsc{GiBUU}, \eHIJING, and \BeAGLE, shows that \textsc{GiBUU} most closely matches the trends of the data. While the \eHIJING~produces similar trends as the data, the values of the widths and the broadenings are smaller than the experimental ones. The functional dependences of the calculated widths with \BeAGLE~match the data the best, which indicates that its mixed hybrid approach might effectively integrate the characteristics of the mostly partonic model \eHIJING~and the mostly hadronic model \textsc{GiBUU}.  While the \BeAGLE~model performs exceptionally well when the observables are integrated over the kinematic phase space, one sees larger deviations from the data when data are binned in multiple variables, such as the secondary-pion transverse momentum.  Overall, the \BeAGLE~calculated values for the widths and broadenings are within $\mathcal{O}(10\%)$ of those of the data.  

These results suggest that quarks or hadrons may deflect by a few hundred mrad due to multiple scatterings in a nucleus. Some models interpret this scattering as reflecting nuclear transport properties or gluon densities, which are anticipated to be independent of the probe or the measured final state. Future experiments at higher energies, like the 11~GeV beam at the Thomas Jefferson National Accelerator Facility (JLab)~\cite{Burkert:2018nvj,Burkert:2020akg,Arrington:2021alx}, or future JLab upgrades~\cite{Accardi:2023chb}, and the upcoming Electron-Ion Colliders in the USA~\cite{Accardi:2012qut,AbdulKhalek:2021gbh} and China~\cite{Anderle:2021wcy}, will further explore these aspects.  A planned experiment with the CLAS12 spectrometer~\cite{Burkert:2020akg} at JLab, in particular, will extend these measurements in more detail, using D, C, Al, Cu, Sn, and Pb targets, at larger luminosity, higher beam energy, and with polarized beams.

\section*{Acknowledgements} \label{sec:acknowledgements}
The authors acknowledge the staff of the Accelerator and Physics Divisions at 
the Thomas Jefferson National Accelerator Facility who made this experiment 
possible. This work was supported in part by the Chilean Agencia Nacional de Investigacion y Desarollo (ANID), by ANID PIA grant 
ACT1413, by ANID PIA/APOYO AFB180002 and AFB230003, by ANID FONDECYT No. 1161642 and No. 1240904 and No. 11181215, by the U.S. Department of Energy, the Italian Instituto Nazionale di Fisica Nucleare, the French Centre 
National de la Recherche Scientifique, the French Commissariat \`a l'Energie 
Atomique, 
the United Kingdom Science and Technology Facilities Council (STFC), the 
Scottish Universities Physics Alliance (SUPA), the National Research Foundation 
of Korea (NRF), the National Science Foundation (NSF),  the HelmholtzForschungsakademie Hessen für FAIR (HFHF), the Ministry of Science and Higher Education of the Russian Federation, and the Office of Research and Economic Development at Mississippi 
State University. This work has received funding from 
the European Research Council (ERC) under the European Union’s Horizon 2020 
research and innovation programme (Grant agreement No. 804480). The Southeastern Universities Research 
Association operated the Thomas Jefferson National Accelerator Facility for the 
United States Department of Energy under Contract No. DE-AC05-06OR23177 at the time this experiment took place.

 \FloatBarrier

\bibliographystyle{utphys} 
\bibliography{biblio.bib} 

\providecommand{\href}[2]{#2}\begingroup\raggedright\begin{thebibliography}{10}

\bibitem{LRP2023}
{Nuclear Science Advisory Committee}, ``{A New Era of Discovery: The 2023 Long Range Plan for Nuclear Science}.''. \url{https://nuclearsciencefuture.org/wp-content/uploads/2023/11/NSAC-LRP-2023-v1.3.pdf}.

\bibitem{Accardi:2009qv}
A.~Accardi, F.~Arleo, W.~K. Brooks, D.~D'Enterria, and V.~Muccifora, ``{Parton propagation and fragmentation in QCD matter},'' \href{http://dx.doi.org/10.1393/ncr/i2009-10048-0}{{\em Riv. Nuovo Cim.} {\bfseries 32} no.~9-10, (2009) 439--554}.

\bibitem{HERMES:2000ytc}
{\bfseries HERMES} Collaboration, A.~Airapetian {\em et~al.}, ``{Hadron formation in deep inelastic positron scattering in a nuclear environment},'' \href{http://dx.doi.org/10.1007/s100520100697}{{\em Eur. Phys. J. C} {\bfseries 20} (2001) 479--486}.

\bibitem{HERMES:2003icw}
{\bfseries HERMES} Collaboration, A.~Airapetian {\em et~al.}, ``{Quark fragmentation to $\pi^\pm$, $\pi^0$, $K^\pm$, $p$ and $\bar p$ in the nuclear environment},'' \href{http://dx.doi.org/10.1016/j.physletb.2003.10.026}{{\em Phys. Lett. B} {\bfseries 577} (2003) 37--46}.

\bibitem{HERMES:2007plz}
{\bfseries HERMES} Collaboration, A.~Airapetian {\em et~al.}, ``{Hadronization in semi-inclusive deep-inelastic scattering on nuclei},'' \href{http://dx.doi.org/10.1016/j.nuclphysb.2007.06.004}{{\em Nucl. Phys. B} {\bfseries 780} (2007) 1--27}.

\bibitem{HERMES:2009uge}
{\bfseries HERMES} Collaboration, A.~Airapetian {\em et~al.}, ``{Transverse momentum broadening of hadrons produced in semi-inclusive deep-inelastic scattering on nuclei},'' \href{http://dx.doi.org/10.1016/j.physletb.2010.01.020}{{\em Phys. Lett. B} {\bfseries 684} (2010) 114--118}.

\bibitem{HERMES:2011qjb}
{\bfseries HERMES} Collaboration, A.~Airapetian {\em et~al.}, ``{Multidimensional study of hadronization in nuclei},'' \href{http://dx.doi.org/10.1140/epja/i2011-11113-5}{{\em Eur. Phys. J. A} {\bfseries 47} (2011) 113}.

\bibitem{CLAS:2011oae}
{\bfseries CLAS} Collaboration, A.~Daniel {\em et~al.}, ``{Measurement of the nuclear multiplicity ratio for $K^0_s$ hadronization at CLAS},'' \href{http://dx.doi.org/10.1016/j.physletb.2011.10.071}{{\em Phys. Lett. B} {\bfseries 706} (2011) 26--31}.

\bibitem{CLAS:2021jhm}
{\bfseries CLAS} Collaboration, S.~Moran {\em et~al.}, ``{Measurement of charged-pion production in deep-inelastic scattering off nuclei with the CLAS detector},'' \href{http://dx.doi.org/10.1103/PhysRevC.105.015201}{{\em Phys. Rev. C} {\bfseries 105} no.~1, (2022) 015201}.

\bibitem{CLAS:2022oux}
{\bfseries CLAS} Collaboration, T.~Chetry {\em et~al.}, ``{First Measurement of \ensuremath{\Lambda} Electroproduction off Nuclei in the Current and Target Fragmentation Regions},'' \href{http://dx.doi.org/10.1103/PhysRevLett.130.142301}{{\em Phys. Rev. Lett.} {\bfseries 130} no.~14, (2023) 142301}.

\bibitem{CLAS:2022asf}
{\bfseries CLAS} Collaboration, S.~J. Paul {\em et~al.}, ``{Observation of Azimuth-Dependent Suppression of Hadron Pairs in Electron Scattering off Nuclei},'' \href{http://dx.doi.org/10.1103/PhysRevLett.129.182501}{{\em Phys. Rev. Lett.} {\bfseries 129} no.~18, (2022) 182501}.

\bibitem{HERMES:2005mar}
{\bfseries HERMES} Collaboration, A.~Airapetian {\em et~al.}, ``{Double Hadron Leptoproduction in the Nuclear Medium},'' \href{http://dx.doi.org/10.1103/PhysRevLett.96.162301}{{\em Phys. Rev. Lett.} {\bfseries 96} (2006) 162301}.

\bibitem{STAR:2021fgw}
{\bfseries STAR} Collaboration, M.~S. Abdallah {\em et~al.}, ``{Evidence for Nonlinear Gluon Effects in QCD and Their Mass Number Dependence at STAR},'' \href{http://dx.doi.org/10.1103/PhysRevLett.129.092501}{{\em Phys. Rev. Lett.} {\bfseries 129} no.~9, (2022) 092501}.

\bibitem{LHCb:2023png}
{\bfseries LHCb} Collaboration, R.~Aaij {\em et~al.}, ``{Measurement of forward charged hadron flow harmonics in peripheral PbPb collisions at $\sqrt{s_{NN}}=5.02$ TeV with the LHCb detector},'' \href{http://dx.doi.org/10.1103/PhysRevC.109.054908}{{\em Phys. Rev. C} {\bfseries 109} (2024) 054908}.

\bibitem{CMS:2012qk}
{\bfseries CMS} Collaboration, S.~Chatrchyan {\em et~al.}, ``{Observation of long-range, near-side angular correlations in pPb collisions at the LHC},'' \href{http://dx.doi.org/10.1016/j.physletb.2012.11.025}{{\em Phys. Lett. B} {\bfseries 718} (2013) 795--814}.

\bibitem{CMS:2015fgy}
{\bfseries CMS} Collaboration, V.~Khachatryan {\em et~al.}, ``{Measurement of Long-Range Near-Side Two-Particle Angular Correlations in $pp$ Collisions at $\sqrt s =$13 TeV},'' \href{http://dx.doi.org/10.1103/PhysRevLett.116.172302}{{\em Phys. Rev. Lett.} {\bfseries 116} no.~17, (2016) 172302}.

\bibitem{Liang:2008vz}
Z.-t. Liang, X.-N. Wang, and J.~Zhou, ``{Transverse-momentum-dependent parton distribution function and jet transport in medium},'' \href{http://dx.doi.org/10.1103/PhysRevD.77.125010}{{\em Phys. Rev. D} {\bfseries 77} (2008) 125010}.

\bibitem{Xing:2012ii}
H.~Xing, Z.-B. Kang, I.~Vitev, and E.~Wang, ``{Transverse momentum imbalance of back-to-back particle production in $p+A$ and $e+A$ collisions},'' \href{http://dx.doi.org/10.1103/PhysRevD.86.094010}{{\em Phys. Rev. D} {\bfseries 86} (2012) 094010}.

\bibitem{Schafer:2013mza}
A.~Sch\"afer and J.~Zhou, ``{Process dependent nuclear $k_\perp$ broadening effect},'' \href{http://dx.doi.org/10.1103/PhysRevD.88.074012}{{\em Phys. Rev. D} {\bfseries 88} no.~7, (2013) 074012}.

\bibitem{Mueller:2016xoc}
A.~H. Mueller, B.~Wu, B.-W. Xiao, and F.~Yuan, ``{Medium induced transverse momentum broadening in hard processes},'' \href{http://dx.doi.org/10.1103/PhysRevD.95.034007}{{\em Phys. Rev. D} {\bfseries 95} no.~3, (2017) 034007}.

\bibitem{Cougoulic:2017ust}
F.~Cougoulic and S.~Peign\'e, ``{Nuclear $p_\perp$-broadening of an energetic parton pair},'' \href{http://dx.doi.org/10.1007/JHEP05(2018)203}{{\em JHEP} {\bfseries 05} (2018) 203}.

\bibitem{Alrashed:2021csd}
M.~Alrashed, D.~Anderle, Z.-B. Kang, J.~Terry, and H.~Xing, ``{Nuclear Modification of Transverse Momentum Dependent Parton Distribution Functions by a Global QCD Analysis},'' \href{http://dx.doi.org/10.1103/PhysRevLett.129.242001}{{\em Phys. Rev. Lett.} {\bfseries 129} no.~24, (2022) 242001}.

\bibitem{Dupre:2023bnt}
R.~Dupr\'e and F.~Ceccopieri, ``{A TMD-based model for Hadronization off heavy nuclei},'' \href{http://arxiv.org/abs/2308.10739}{{\ttfamily arXiv:2308.10739 [hep-ph]}}.

\bibitem{CLAS:2003umf}
{\bfseries CLAS} Collaboration, B.~A. Mecking {\em et~al.}, ``{The CEBAF large acceptance spectrometer (CLAS)},'' \href{http://dx.doi.org/10.1016/S0168-9002(03)01001-5}{{\em Nucl. Instrum. Meth. A} {\bfseries 503} (2003) 513--553}.

\bibitem{Hakobyan:2008zz}
H.~Hakobyan {\em et~al.}, ``{A double-target system for precision measurements of nuclear medium effects},'' \href{http://dx.doi.org/10.1016/j.nima.2008.04.055}{{\em Nucl. Instrum. Meth. A} {\bfseries 592} (2008) 218--223}.

\bibitem{Sjostrand:2003wg}
T.~Sjostrand, L.~Lonnblad, S.~Mrenna, and P.~Z. Skands, ``{PYTHIA 6.3 Physics and Manual},'' \href{http://arxiv.org/abs/hep-ph/0308153}{{\ttfamily arXiv:hep-ph/0308153}}.

\bibitem{GSIM}
E.~Wolin, ``{GSIM User’s Guide Version CERN 1.0},'' 1995.
\newblock \url{https://www.jlab.org/Hall-B/document/gsim/userguide.html}.

\bibitem{Brun:1994aa}
R.~Brun, F.~Bruyant, F.~Carminati, S.~Giani, M.~Maire, A.~McPherson, G.~Patrick, and L.~Urban, ``{GEANT Detector Description and Simulation Tool},''.
\url{https://doi.org/10.17181/CERN.MUHF.DMJ1}.

\bibitem{Buss:2011mx}
O.~Buss, T.~Gaitanos, K.~Gallmeister, H.~van Hees, M.~Kaskulov, O.~Lalakulich, A.~B. Larionov, T.~Leitner, J.~Weil, and U.~Mosel, ``{Transport-theoretical description of nuclear reactions},'' \href{http://dx.doi.org/10.1016/j.physrep.2011.12.001}{{\em Phys. Rept.} {\bfseries 512} (2012) 1--124}.

\bibitem{ElAlaoui_PC}
A.~El~Alaoui.
\newblock Private communication. In this tuning, based on proton-target data in the context of Ref.~\cite{CLAS:2022sqt}, the sigma of the intrinsic transverse-momentum distribution of quarks within a nucleon, $k_\perp$, was changed from 0.44 GeV to 0.64 GeV. The sigma of the distribution of transverse momentum, $p_\perp$, of hadrons produced in the LUND string breaking was changed from 0.36 GeV to 0.5 GeV. The value of the ratio of light vector mesons to total light mesons produced in the fragmentation was changed from 0.5 to 0.7. The value of the ratio of strange vector mesons to total strange mesons produced in the fragmentation was changed from 0.6 to 0.75.

\bibitem{CLAS:2022sqt}
{\bfseries CLAS} Collaboration, H.~Avakian {\em et~al.}, ``{Observation of Correlations between Spin and Transverse Momenta in Back-to-Back Dihadron Production at CLAS12},'' \href{http://dx.doi.org/10.1103/PhysRevLett.130.022501}{{\em Phys. Rev. Lett.} {\bfseries 130} no.~2, (2023) 022501}.

\bibitem{Ke:2023xeo}
W.~Ke, Y.-Y. Zhang, H.~Xing, and X.-N. Wang, ``{eHIJING: an Event Generator for Jet Tomography in Electron-Ion Collisions},'' \href{http://arxiv.org/abs/2304.10779}{{\ttfamily arXiv:2304.10779 [hep-ph]}}.

\bibitem{Chang:2022hkt}
W.~Chang, E.-C. Aschenauer, M.~D. Baker, A.~Jentsch, J.-H. Lee, Z.~Tu, Z.~Yin, and L.~Zheng, ``{Benchmark $eA$ generator for leptoproduction in high-energy lepton-nucleus collisions},'' \href{http://dx.doi.org/10.1103/PhysRevD.106.012007}{{\em Phys. Rev. D} {\bfseries 106} no.~1, (2022) 012007}.

\bibitem{bierlich2022comprehensive}
C.~Bierlich, S.~Chakraborty, N.~Desai, L.~Gellersen, I.~Helenius, P.~Ilten, L.~Lönnblad, S.~Mrenna, S.~Prestel, C.~T. Preuss, T.~Sjöstrand, P.~Skands, M.~Utheim, and R.~Verheyen, ``{A comprehensive guide to the physics and usage of PYTHIA 8.3},'' \href{http://arxiv.org/abs/2203.11601}{{\ttfamily arXiv:2203.11601 [hep-ph]}}.

\bibitem{Sjostrand:2006za}
T.~Sjostrand, S.~Mrenna, and P.~Z. Skands, ``{PYTHIA 6.4 Physics and Manual},'' \href{http://dx.doi.org/10.1088/1126-6708/2006/05/026}{{\em JHEP} {\bfseries 05} (2006) 026}.

\bibitem{Whalley:2005nh}
M.~R. Whalley, D.~Bourilkov, and R.~C. Group, ``{The Les Houches accord PDFs (LHAPDF) and LHAGLUE},'' in {\em {HERA and the LHC: A Workshop on the Implications of HERA and LHC Physics (Startup Meeting, CERN, 26-27 March 2004; Midterm Meeting, CERN, 11-13 October 2004)}}, pp.~575--581.
\newblock 8, 2005.

\bibitem{Roesler:2000he}
S.~Roesler, R.~Engel, and J.~Ranft, \href{http://dx.doi.org/10.1007/978-3-642-18211-2_166}{``{The Monte Carlo Event Generator DPMJET-III},''} in {\em {International Conference on Advanced Monte Carlo for Radiation Physics, Particle Transport Simulation and Applications (MC 2000)}}, pp.~1033--1038.
\newblock 12, 2000.

\bibitem{Dupre:2011afa}
R.~Dupr\'e, {\em {Quark Fragmentation and Hadron Formation in Nuclear Matter}}.
\newblock PhD thesis, Lyon, IPN, 2011.
\newblock \url{https://theses.hal.science/tel-00751424/document}.

\bibitem{Ferrari:2005zk}
A.~Ferrari, P.~R. Sala, A.~Fasso, and J.~Ranft, ``{FLUKA: A Multi-Particle Transport Code},''. \url{https://doi.org/10.2172/877507}. (Program version 2005).

\bibitem{Burkert:2018nvj}
{\bfseries CLAS} Collaboration, V.~D. Burkert, ``{Jefferson Lab at 12 GeV: The Science Program},'' \href{http://dx.doi.org/10.1146/annurev-nucl-101917-021129}{{\em Ann. Rev. Nucl. Part. Sci.} {\bfseries 68} (2018) 405--428}.

\bibitem{Burkert:2020akg}
V.~D. Burkert {\em et~al.}, ``{The CLAS12 Spectrometer at Jefferson Laboratory},'' \href{http://dx.doi.org/10.1016/j.nima.2020.163419}{{\em Nucl. Instrum. Meth. A} {\bfseries 959} (2020) 163419}.

\bibitem{Arrington:2021alx}
J.~Arrington {\em et~al.}, ``{Physics with CEBAF at 12 GeV and future opportunities},'' \href{http://dx.doi.org/https://doi.org/10.1016/j.ppnp.2022.103985}{{\em Progress in Particle and Nuclear Physics} {\bfseries 127} (2022) 103985}.

\bibitem{Accardi:2023chb}
A.~Accardi {\em et~al.}, ``{Strong Interaction Physics at the Luminosity Frontier with 22 GeV Electrons at Jefferson Lab},'' \href{http://arxiv.org/abs/2306.09360}{{\ttfamily arXiv:2306.09360 [nucl-ex]}}. Accepted for publication in Eur. Phys. J. A.

\bibitem{Accardi:2012qut}
A.~Accardi {\em et~al.}, ``{Electron-Ion Collider: The next QCD frontier},''
\href{http://dx.doi.org/10.1140/epja/i2016-16268-9}{{\em Eur. Phys. J.} {\bfseries A52} no.~9, (2016) 268}.

\bibitem{AbdulKhalek:2021gbh}
R.~Abdul~Khalek {\em et~al.}, ``{Science Requirements and Detector Concepts for the Electron-Ion Collider: EIC Yellow Report},'' \href{http://dx.doi.org/https://doi.org/10.1016/j.nuclphysa.2022.122447}{{\em Nuclear Physics A} {\bfseries 1026} (2022) 122447}.

\bibitem{Anderle:2021wcy}
D.~P. Anderle {\em et~al.}, ``{Electron-ion collider in China},'' \href{http://dx.doi.org/10.1007/s11467-021-1062-0}{{\em Front. Phys. (Beijing)} {\bfseries 16} no.~6, (2021) 64701}.

\end{thebibliography}\endgroup
\appendix
\section{Tabulations of Our Measurements}
\label{app:tables}
We tabulate the measurements for the correlations in Tables~\ref{tab:all_data}-\ref{tab:pT2_data}, the ratios, $R$, in Tables~\ref{tab:all_R_data}--\ref{tab:pT2_R_data}, the widths in Table~\ref{tab:width_data}, and the broadening in Table~\ref{tab:broadening_data}.
In each of these tables, the first uncertainty is statistical, and the second is systematic.
\setlength{\tabcolsep}{6pt}
\begin{table*}[]
    \centering
    \begin{tabular}{c|c|c|c|c|c}
         & $\Delta\phi$ [rad] & D & C & Fe & Pb \\
         \hline
          & $0.00-0.39$ & 0.029$\pm$0.001$\pm$0.004 & 0.038$\pm$0.002$\pm$0.006 & 0.039$\pm$0.002$\pm$0.007 & 0.039$\pm$0.003$\pm$0.008 \\
 & $0.39-0.79$ & 0.036$\pm$0.001$\pm$0.004 & 0.045$\pm$0.002$\pm$0.005 & 0.043$\pm$0.002$\pm$0.005 & 0.039$\pm$0.003$\pm$0.005 \\
 & $0.79-1.18$ & 0.053$\pm$0.002$\pm$0.004 & 0.066$\pm$0.003$\pm$0.006 & 0.056$\pm$0.002$\pm$0.005 & 0.063$\pm$0.004$\pm$0.005 \\
 & $1.18-1.57$ & 0.085$\pm$0.002$\pm$0.006 & 0.096$\pm$0.003$\pm$0.007 & 0.081$\pm$0.003$\pm$0.006 & 0.085$\pm$0.005$\pm$0.006 \\
 & $1.57-1.96$ & 0.152$\pm$0.003$\pm$0.010 & 0.146$\pm$0.004$\pm$0.010 & 0.122$\pm$0.003$\pm$0.008 & 0.132$\pm$0.006$\pm$0.009 \\
 & $1.96-2.36$ & 0.233$\pm$0.003$\pm$0.015 & 0.205$\pm$0.004$\pm$0.013 & 0.164$\pm$0.003$\pm$0.010 & 0.158$\pm$0.006$\pm$0.010 \\
 & $2.36-2.75$ & 0.344$\pm$0.004$\pm$0.022 & 0.281$\pm$0.005$\pm$0.018 & 0.223$\pm$0.004$\pm$0.015 & 0.207$\pm$0.007$\pm$0.014 \\
 & $2.75-3.14$ & 0.394$\pm$0.004$\pm$0.030 & 0.302$\pm$0.005$\pm$0.024 & 0.245$\pm$0.004$\pm$0.020 & 0.227$\pm$0.007$\pm$0.019 \\
\hline

    \end{tabular}
    \caption{Tabulation of the dependence of $C$ on $|\Delta\phi|$ integrated over all other kinematic variables.  These values are shown in the top panel of Fig.~\ref{fig:correlations}.}
    \label{tab:all_data}
\end{table*}

\begin{table*}[]
    \centering
    \begin{tabular}{c|c|c|c|c|c}
        $\Delta Y$ & $\Delta\phi$ [rad] & D & C & Fe & Pb \\
         \hline
         $-0.50-0.50$ & $0.00-0.39$ & 0.009$\pm$0.002$\pm$0.001 & 0.014$\pm$0.004$\pm$0.003 & 0.024$\pm$0.004$\pm$0.005 & 0.010$\pm$0.005$\pm$0.002 \\
 & $0.39-0.79$ & 0.009$\pm$0.002$\pm$0.001 & 0.024$\pm$0.005$\pm$0.003 & 0.026$\pm$0.004$\pm$0.004 & 0.026$\pm$0.008$\pm$0.004 \\
 & $0.79-1.18$ & 0.018$\pm$0.003$\pm$0.002 & 0.034$\pm$0.006$\pm$0.004 & 0.029$\pm$0.005$\pm$0.003 & 0.056$\pm$0.011$\pm$0.007 \\
 & $1.18-1.57$ & 0.066$\pm$0.005$\pm$0.005 & 0.079$\pm$0.009$\pm$0.006 & 0.082$\pm$0.008$\pm$0.006 & 0.091$\pm$0.014$\pm$0.007 \\
 & $1.57-1.96$ & 0.118$\pm$0.007$\pm$0.011 & 0.148$\pm$0.011$\pm$0.013 & 0.118$\pm$0.009$\pm$0.011 & 0.125$\pm$0.016$\pm$0.011 \\
 & $1.96-2.36$ & 0.247$\pm$0.010$\pm$0.016 & 0.231$\pm$0.013$\pm$0.015 & 0.199$\pm$0.011$\pm$0.013 & 0.162$\pm$0.017$\pm$0.011 \\
 & $2.36-2.75$ & 0.407$\pm$0.012$\pm$0.026 & 0.345$\pm$0.016$\pm$0.022 & 0.297$\pm$0.014$\pm$0.019 & 0.277$\pm$0.022$\pm$0.018 \\
 & $2.75-3.14$ & 0.493$\pm$0.014$\pm$0.037 & 0.371$\pm$0.018$\pm$0.028 & 0.317$\pm$0.015$\pm$0.025 & 0.295$\pm$0.024$\pm$0.024 \\
\hline
$0.50-1.50$ & $0.00-0.39$ & 0.024$\pm$0.001$\pm$0.004 & 0.035$\pm$0.002$\pm$0.009 & 0.035$\pm$0.002$\pm$0.011 & 0.036$\pm$0.004$\pm$0.013 \\
 & $0.39-0.79$ & 0.030$\pm$0.002$\pm$0.003 & 0.043$\pm$0.003$\pm$0.006 & 0.041$\pm$0.002$\pm$0.006 & 0.032$\pm$0.004$\pm$0.005 \\
 & $0.79-1.18$ & 0.048$\pm$0.002$\pm$0.004 & 0.064$\pm$0.003$\pm$0.005 & 0.057$\pm$0.003$\pm$0.004 & 0.058$\pm$0.005$\pm$0.004 \\
 & $1.18-1.57$ & 0.084$\pm$0.003$\pm$0.006 & 0.098$\pm$0.004$\pm$0.007 & 0.081$\pm$0.003$\pm$0.006 & 0.082$\pm$0.006$\pm$0.006 \\
 & $1.57-1.96$ & 0.159$\pm$0.003$\pm$0.011 & 0.147$\pm$0.005$\pm$0.010 & 0.123$\pm$0.004$\pm$0.009 & 0.135$\pm$0.007$\pm$0.010 \\
 & $1.96-2.36$ & 0.237$\pm$0.004$\pm$0.015 & 0.207$\pm$0.005$\pm$0.013 & 0.165$\pm$0.004$\pm$0.010 & 0.165$\pm$0.007$\pm$0.010 \\
 & $2.36-2.75$ & 0.348$\pm$0.005$\pm$0.022 & 0.283$\pm$0.006$\pm$0.018 & 0.221$\pm$0.005$\pm$0.014 & 0.210$\pm$0.008$\pm$0.014 \\
 & $2.75-3.14$ & 0.398$\pm$0.005$\pm$0.029 & 0.299$\pm$0.007$\pm$0.022 & 0.241$\pm$0.005$\pm$0.019 & 0.228$\pm$0.009$\pm$0.019 \\
\hline
$1.50-2.50$ & $0.00-0.39$ & 0.054$\pm$0.003$\pm$0.005 & 0.057$\pm$0.005$\pm$0.005 & 0.056$\pm$0.004$\pm$0.005 & 0.064$\pm$0.008$\pm$0.006 \\
 & $0.39-0.79$ & 0.064$\pm$0.004$\pm$0.006 & 0.060$\pm$0.005$\pm$0.005 & 0.058$\pm$0.005$\pm$0.005 & 0.066$\pm$0.008$\pm$0.006 \\
 & $0.79-1.18$ & 0.084$\pm$0.004$\pm$0.008 & 0.087$\pm$0.006$\pm$0.009 & 0.067$\pm$0.005$\pm$0.007 & 0.082$\pm$0.009$\pm$0.008 \\
 & $1.18-1.57$ & 0.099$\pm$0.004$\pm$0.008 & 0.099$\pm$0.006$\pm$0.008 & 0.079$\pm$0.005$\pm$0.007 & 0.087$\pm$0.009$\pm$0.007 \\
 & $1.57-1.96$ & 0.150$\pm$0.005$\pm$0.010 & 0.144$\pm$0.008$\pm$0.009 & 0.120$\pm$0.006$\pm$0.008 & 0.126$\pm$0.011$\pm$0.008 \\
 & $1.96-2.36$ & 0.216$\pm$0.006$\pm$0.014 & 0.186$\pm$0.008$\pm$0.012 & 0.146$\pm$0.007$\pm$0.009 & 0.137$\pm$0.011$\pm$0.009 \\
 & $2.36-2.75$ & 0.302$\pm$0.007$\pm$0.022 & 0.244$\pm$0.010$\pm$0.018 & 0.191$\pm$0.008$\pm$0.014 & 0.167$\pm$0.012$\pm$0.012 \\
 & $2.75-3.14$ & 0.340$\pm$0.008$\pm$0.029 & 0.276$\pm$0.010$\pm$0.025 & 0.222$\pm$0.008$\pm$0.020 & 0.193$\pm$0.013$\pm$0.018 \\
\hline

    \end{tabular}
    \caption{Same as Table~\ref{tab:all_data}, except in slices of $\Delta Y$.  These values are shown in the top row of Fig.~\ref{fig:dY}.}
    \label{tab:dy_data}
\end{table*}

\begin{table*}[]
    \centering
    \begin{tabular}{c|c|c|c|c|c}
        $p^T_1$ [GeV] & $\Delta\phi$ [rad] & D & C & Fe & Pb \\
         \hline
         $0.25-0.40$ & $0.00-0.39$ & 0.059$\pm$0.003$\pm$0.007 & 0.064$\pm$0.005$\pm$0.008 & 0.063$\pm$0.005$\pm$0.008 & 0.073$\pm$0.009$\pm$0.010 \\
 & $0.39-0.79$ & 0.074$\pm$0.004$\pm$0.007 & 0.074$\pm$0.005$\pm$0.007 & 0.072$\pm$0.005$\pm$0.007 & 0.069$\pm$0.008$\pm$0.007 \\
 & $0.79-1.18$ & 0.091$\pm$0.004$\pm$0.008 & 0.097$\pm$0.006$\pm$0.008 & 0.081$\pm$0.005$\pm$0.007 & 0.091$\pm$0.009$\pm$0.008 \\
 & $1.18-1.57$ & 0.117$\pm$0.004$\pm$0.008 & 0.124$\pm$0.007$\pm$0.008 & 0.098$\pm$0.006$\pm$0.006 & 0.106$\pm$0.010$\pm$0.007 \\
 & $1.57-1.96$ & 0.208$\pm$0.006$\pm$0.013 & 0.181$\pm$0.008$\pm$0.012 & 0.152$\pm$0.007$\pm$0.010 & 0.157$\pm$0.012$\pm$0.010 \\
 & $1.96-2.36$ & 0.242$\pm$0.006$\pm$0.015 & 0.209$\pm$0.008$\pm$0.013 & 0.175$\pm$0.007$\pm$0.011 & 0.184$\pm$0.013$\pm$0.012 \\
 & $2.36-2.75$ & 0.262$\pm$0.006$\pm$0.017 & 0.217$\pm$0.009$\pm$0.014 & 0.190$\pm$0.007$\pm$0.013 & 0.149$\pm$0.011$\pm$0.010 \\
 & $2.75-3.14$ & 0.273$\pm$0.006$\pm$0.023 & 0.240$\pm$0.009$\pm$0.022 & 0.209$\pm$0.008$\pm$0.019 & 0.172$\pm$0.012$\pm$0.016 \\
\hline
$0.40-0.60$ & $0.00-0.39$ & 0.022$\pm$0.002$\pm$0.003 & 0.033$\pm$0.003$\pm$0.005 & 0.035$\pm$0.003$\pm$0.006 & 0.035$\pm$0.005$\pm$0.008 \\
 & $0.39-0.79$ & 0.028$\pm$0.002$\pm$0.002 & 0.044$\pm$0.003$\pm$0.004 & 0.039$\pm$0.003$\pm$0.004 & 0.038$\pm$0.005$\pm$0.004 \\
 & $0.79-1.18$ & 0.048$\pm$0.002$\pm$0.003 & 0.064$\pm$0.004$\pm$0.004 & 0.050$\pm$0.003$\pm$0.003 & 0.059$\pm$0.006$\pm$0.004 \\
 & $1.18-1.57$ & 0.084$\pm$0.003$\pm$0.005 & 0.093$\pm$0.005$\pm$0.006 & 0.079$\pm$0.004$\pm$0.005 & 0.089$\pm$0.007$\pm$0.006 \\
 & $1.57-1.96$ & 0.159$\pm$0.004$\pm$0.011 & 0.152$\pm$0.006$\pm$0.010 & 0.124$\pm$0.005$\pm$0.009 & 0.150$\pm$0.009$\pm$0.010 \\
 & $1.96-2.36$ & 0.253$\pm$0.005$\pm$0.016 & 0.214$\pm$0.007$\pm$0.013 & 0.170$\pm$0.005$\pm$0.011 & 0.159$\pm$0.009$\pm$0.010 \\
 & $2.36-2.75$ & 0.346$\pm$0.006$\pm$0.022 & 0.264$\pm$0.008$\pm$0.017 & 0.213$\pm$0.006$\pm$0.014 & 0.201$\pm$0.010$\pm$0.013 \\
 & $2.75-3.14$ & 0.384$\pm$0.006$\pm$0.027 & 0.299$\pm$0.008$\pm$0.022 & 0.237$\pm$0.007$\pm$0.018 & 0.230$\pm$0.011$\pm$0.018 \\
\hline
$0.60-1.00$ & $0.00-0.39$ & 0.006$\pm$0.001$\pm$0.001 & 0.020$\pm$0.003$\pm$0.005 & 0.023$\pm$0.003$\pm$0.006 & 0.019$\pm$0.004$\pm$0.005 \\
 & $0.39-0.79$ & 0.007$\pm$0.001$\pm$0.001 & 0.023$\pm$0.003$\pm$0.002 & 0.025$\pm$0.003$\pm$0.002 & 0.021$\pm$0.004$\pm$0.002 \\
 & $0.79-1.18$ & 0.019$\pm$0.002$\pm$0.001 & 0.039$\pm$0.004$\pm$0.003 & 0.043$\pm$0.003$\pm$0.003 & 0.047$\pm$0.006$\pm$0.004 \\
 & $1.18-1.57$ & 0.052$\pm$0.003$\pm$0.004 & 0.078$\pm$0.005$\pm$0.006 & 0.070$\pm$0.004$\pm$0.005 & 0.067$\pm$0.007$\pm$0.005 \\
 & $1.57-1.96$ & 0.083$\pm$0.004$\pm$0.006 & 0.114$\pm$0.006$\pm$0.008 & 0.099$\pm$0.005$\pm$0.007 & 0.101$\pm$0.009$\pm$0.007 \\
 & $1.96-2.36$ & 0.193$\pm$0.006$\pm$0.012 & 0.192$\pm$0.008$\pm$0.013 & 0.156$\pm$0.006$\pm$0.010 & 0.147$\pm$0.010$\pm$0.010 \\
 & $2.36-2.75$ & 0.431$\pm$0.008$\pm$0.027 & 0.371$\pm$0.011$\pm$0.023 & 0.271$\pm$0.008$\pm$0.017 & 0.278$\pm$0.014$\pm$0.018 \\
 & $2.75-3.14$ & 0.534$\pm$0.010$\pm$0.036 & 0.361$\pm$0.011$\pm$0.025 & 0.293$\pm$0.009$\pm$0.021 & 0.271$\pm$0.014$\pm$0.021 \\
\hline

    \end{tabular}
    \caption{Same as Table~\ref{tab:all_data}, except in slices of $p^T_1$. These values are shown in the top row of Fig.~\ref{fig:pT1}.}
    \label{tab:pT1_data}
\end{table*}

\begin{table*}[]
    \centering
         
    \begin{tabular}{c|c|c|c|c|c}
         $p^T_1$ [GeV]& $\Delta\phi$ [rad] & D & C & Fe & Pb \\
         \hline
         $0.25-0.40$ & $0.00-0.39$ & 0.039$\pm$0.002$\pm$0.004 & 0.040$\pm$0.003$\pm$0.006 & 0.046$\pm$0.003$\pm$0.007 & 0.046$\pm$0.005$\pm$0.009 \\
 & $0.39-0.79$ & 0.049$\pm$0.002$\pm$0.003 & 0.052$\pm$0.003$\pm$0.004 & 0.051$\pm$0.003$\pm$0.004 & 0.047$\pm$0.005$\pm$0.004 \\
 & $0.79-1.18$ & 0.065$\pm$0.002$\pm$0.004 & 0.076$\pm$0.004$\pm$0.005 & 0.065$\pm$0.003$\pm$0.004 & 0.068$\pm$0.005$\pm$0.004 \\
 & $1.18-1.57$ & 0.104$\pm$0.003$\pm$0.007 & 0.107$\pm$0.005$\pm$0.007 & 0.092$\pm$0.004$\pm$0.006 & 0.097$\pm$0.007$\pm$0.006 \\
 & $1.57-1.96$ & 0.178$\pm$0.004$\pm$0.011 & 0.153$\pm$0.005$\pm$0.010 & 0.130$\pm$0.004$\pm$0.008 & 0.145$\pm$0.008$\pm$0.009 \\
 & $1.96-2.36$ & 0.246$\pm$0.004$\pm$0.015 & 0.197$\pm$0.006$\pm$0.012 & 0.160$\pm$0.005$\pm$0.010 & 0.157$\pm$0.008$\pm$0.010 \\
 & $2.36-2.75$ & 0.315$\pm$0.005$\pm$0.020 & 0.237$\pm$0.006$\pm$0.015 & 0.194$\pm$0.005$\pm$0.013 & 0.189$\pm$0.009$\pm$0.013 \\
 & $2.75-3.14$ & 0.334$\pm$0.005$\pm$0.025 & 0.236$\pm$0.006$\pm$0.018 & 0.200$\pm$0.005$\pm$0.017 & 0.185$\pm$0.009$\pm$0.018 \\
\hline
$0.40-0.60$ & $0.00-0.39$ & 0.018$\pm$0.002$\pm$0.002 & 0.037$\pm$0.003$\pm$0.007 & 0.031$\pm$0.003$\pm$0.007 & 0.030$\pm$0.004$\pm$0.007 \\
 & $0.39-0.79$ & 0.022$\pm$0.002$\pm$0.002 & 0.038$\pm$0.003$\pm$0.004 & 0.036$\pm$0.003$\pm$0.004 & 0.030$\pm$0.004$\pm$0.004 \\
 & $0.79-1.18$ & 0.043$\pm$0.002$\pm$0.003 & 0.056$\pm$0.004$\pm$0.004 & 0.049$\pm$0.003$\pm$0.003 & 0.062$\pm$0.006$\pm$0.004 \\
 & $1.18-1.57$ & 0.068$\pm$0.003$\pm$0.004 & 0.088$\pm$0.005$\pm$0.005 & 0.071$\pm$0.004$\pm$0.004 & 0.074$\pm$0.007$\pm$0.005 \\
 & $1.57-1.96$ & 0.133$\pm$0.004$\pm$0.008 & 0.141$\pm$0.006$\pm$0.009 & 0.116$\pm$0.005$\pm$0.007 & 0.118$\pm$0.009$\pm$0.007 \\
 & $1.96-2.36$ & 0.229$\pm$0.005$\pm$0.014 & 0.214$\pm$0.007$\pm$0.013 & 0.166$\pm$0.006$\pm$0.011 & 0.159$\pm$0.009$\pm$0.010 \\
 & $2.36-2.75$ & 0.362$\pm$0.006$\pm$0.023 & 0.312$\pm$0.009$\pm$0.020 & 0.242$\pm$0.007$\pm$0.015 & 0.211$\pm$0.011$\pm$0.013 \\
 & $2.75-3.14$ & 0.451$\pm$0.007$\pm$0.029 & 0.360$\pm$0.009$\pm$0.024 & 0.283$\pm$0.008$\pm$0.019 & 0.260$\pm$0.012$\pm$0.018 \\
\hline
$0.60-0.80$ & $0.00-0.39$ & 0.006$\pm$0.002$\pm$0.000 & 0.013$\pm$0.005$\pm$0.001 & 0.018$\pm$0.005$\pm$0.001 & 0.023$\pm$0.009$\pm$0.001 \\
 & $0.39-0.79$ & 0.005$\pm$0.002$\pm$0.000 & 0.030$\pm$0.007$\pm$0.002 & 0.020$\pm$0.005$\pm$0.001 & 0.029$\pm$0.011$\pm$0.002 \\
 & $0.79-1.18$ & 0.012$\pm$0.003$\pm$0.001 & 0.038$\pm$0.008$\pm$0.002 & 0.016$\pm$0.004$\pm$0.001 & 0.027$\pm$0.010$\pm$0.002 \\
 & $1.18-1.57$ & 0.018$\pm$0.004$\pm$0.001 & 0.048$\pm$0.009$\pm$0.003 & 0.042$\pm$0.008$\pm$0.003 & 0.044$\pm$0.013$\pm$0.003 \\
 & $1.57-1.96$ & 0.046$\pm$0.006$\pm$0.003 & 0.110$\pm$0.013$\pm$0.007 & 0.076$\pm$0.010$\pm$0.005 & 0.115$\pm$0.020$\pm$0.007 \\
 & $1.96-2.36$ & 0.160$\pm$0.011$\pm$0.010 & 0.213$\pm$0.018$\pm$0.013 & 0.179$\pm$0.015$\pm$0.011 & 0.162$\pm$0.024$\pm$0.010 \\
 & $2.36-2.75$ & 0.492$\pm$0.018$\pm$0.031 & 0.466$\pm$0.026$\pm$0.029 & 0.352$\pm$0.020$\pm$0.022 & 0.335$\pm$0.034$\pm$0.021 \\
 & $2.75-3.14$ & 0.553$\pm$0.021$\pm$0.040 & 0.474$\pm$0.028$\pm$0.034 & 0.380$\pm$0.022$\pm$0.027 & 0.367$\pm$0.038$\pm$0.027 \\
\hline

    \end{tabular}
    \caption{Same as Table~\ref{tab:all_data}, except in slices of $p^T_2$. These values are shown in the top row of Fig.~\ref{fig:pT2}.}
    \label{tab:pT2_data}
\end{table*}

\begin{table*}[]
    \centering
    \begin{tabular}{c|c | c|c|c}
         & $\Delta\phi$ [rad] & C & Fe & Pb \\
         \hline
          & $0.00-0.39$ & 1.296$\pm$0.081$\pm$0.094 & 1.336$\pm$0.078$\pm$0.123 & 1.354$\pm$0.114$\pm$0.155 \\
 & $0.39-0.79$ & 1.264$\pm$0.072$\pm$0.060 & 1.207$\pm$0.065$\pm$0.061 & 1.102$\pm$0.090$\pm$0.064 \\
 & $0.79-1.18$ & 1.238$\pm$0.060$\pm$0.037 & 1.044$\pm$0.050$\pm$0.031 & 1.184$\pm$0.080$\pm$0.036 \\
 & $1.18-1.57$ & 1.129$\pm$0.043$\pm$0.024 & 0.948$\pm$0.036$\pm$0.020 & 0.992$\pm$0.055$\pm$0.022 \\
 & $1.57-1.96$ & 0.960$\pm$0.030$\pm$0.023 & 0.800$\pm$0.025$\pm$0.019 & 0.869$\pm$0.040$\pm$0.021 \\
 & $1.96-2.36$ & 0.878$\pm$0.024$\pm$0.019 & 0.703$\pm$0.019$\pm$0.015 & 0.676$\pm$0.029$\pm$0.015 \\
 & $2.36-2.75$ & 0.818$\pm$0.019$\pm$0.018 & 0.647$\pm$0.015$\pm$0.014 & 0.602$\pm$0.022$\pm$0.014 \\
 & $2.75-3.14$ & 0.765$\pm$0.016$\pm$0.017 & 0.621$\pm$0.013$\pm$0.015 & 0.576$\pm$0.020$\pm$0.017 \\
\hline

    \end{tabular}
    \caption{Tabulation of the dependence of the ratios, $R$, on $|\Delta\phi|$ integrated over all other kinematic variables. These values are shown in the the bottom row of Fig.~\ref{fig:correlations}.}
    \label{tab:all_R_data}
\end{table*}

\begin{table*}[]
    \centering
    \begin{tabular}{c|c|c|c|c}
        $\Delta Y$ & $\Delta\phi$ [rad] & C & Fe & Pb \\
         \hline
         $-0.50-0.50$ & $0.00-0.39$ & 1.556$\pm$0.479$\pm$0.236 & 2.575$\pm$0.660$\pm$0.471 & 1.077$\pm$0.517$\pm$0.183 \\
 & $0.39-0.79$ & 2.589$\pm$0.684$\pm$0.559 & 2.726$\pm$0.679$\pm$0.588 & 2.731$\pm$0.908$\pm$0.589 \\
 & $0.79-1.18$ & 1.876$\pm$0.402$\pm$0.133 & 1.587$\pm$0.336$\pm$0.112 & 3.136$\pm$0.744$\pm$0.222 \\
 & $1.18-1.57$ & 1.195$\pm$0.153$\pm$0.066 & 1.241$\pm$0.147$\pm$0.068 & 1.390$\pm$0.231$\pm$0.077 \\
 & $1.57-1.96$ & 1.254$\pm$0.120$\pm$0.041 & 0.995$\pm$0.096$\pm$0.032 & 1.057$\pm$0.149$\pm$0.034 \\
 & $1.96-2.36$ & 0.934$\pm$0.073$\pm$0.023 & 0.805$\pm$0.062$\pm$0.020 & 0.657$\pm$0.084$\pm$0.016 \\
 & $2.36-2.75$ & 0.849$\pm$0.054$\pm$0.025 & 0.730$\pm$0.045$\pm$0.022 & 0.680$\pm$0.067$\pm$0.020 \\
 & $2.75-3.14$ & 0.753$\pm$0.043$\pm$0.017 & 0.642$\pm$0.036$\pm$0.016 & 0.598$\pm$0.054$\pm$0.017 \\
\hline
$0.50-1.50$ & $0.00-0.39$ & 1.479$\pm$0.123$\pm$0.174 & 1.485$\pm$0.116$\pm$0.232 & 1.511$\pm$0.169$\pm$0.335 \\
 & $0.39-0.79$ & 1.456$\pm$0.108$\pm$0.073 & 1.367$\pm$0.097$\pm$0.081 & 1.070$\pm$0.122$\pm$0.084 \\
 & $0.79-1.18$ & 1.325$\pm$0.083$\pm$0.034 & 1.168$\pm$0.071$\pm$0.031 & 1.193$\pm$0.106$\pm$0.035 \\
 & $1.18-1.57$ & 1.177$\pm$0.056$\pm$0.025 & 0.972$\pm$0.046$\pm$0.021 & 0.980$\pm$0.069$\pm$0.022 \\
 & $1.57-1.96$ & 0.920$\pm$0.036$\pm$0.020 & 0.775$\pm$0.030$\pm$0.017 & 0.851$\pm$0.049$\pm$0.019 \\
 & $1.96-2.36$ & 0.873$\pm$0.030$\pm$0.021 & 0.695$\pm$0.024$\pm$0.016 & 0.697$\pm$0.037$\pm$0.017 \\
 & $2.36-2.75$ & 0.814$\pm$0.023$\pm$0.018 & 0.634$\pm$0.018$\pm$0.014 & 0.603$\pm$0.028$\pm$0.014 \\
 & $2.75-3.14$ & 0.753$\pm$0.020$\pm$0.017 & 0.607$\pm$0.016$\pm$0.015 & 0.574$\pm$0.025$\pm$0.017 \\
\hline
$1.50-2.50$ & $0.00-0.39$ & 1.060$\pm$0.107$\pm$0.032 & 1.053$\pm$0.099$\pm$0.032 & 1.194$\pm$0.158$\pm$0.036 \\
 & $0.39-0.79$ & 0.944$\pm$0.089$\pm$0.054 & 0.912$\pm$0.081$\pm$0.052 & 1.031$\pm$0.131$\pm$0.059 \\
 & $0.79-1.18$ & 1.035$\pm$0.087$\pm$0.059 & 0.790$\pm$0.068$\pm$0.045 & 0.967$\pm$0.116$\pm$0.055 \\
 & $1.18-1.57$ & 1.002$\pm$0.076$\pm$0.029 & 0.802$\pm$0.061$\pm$0.023 & 0.884$\pm$0.099$\pm$0.027 \\
 & $1.57-1.96$ & 0.957$\pm$0.061$\pm$0.036 & 0.801$\pm$0.050$\pm$0.031 & 0.837$\pm$0.079$\pm$0.032 \\
 & $1.96-2.36$ & 0.860$\pm$0.049$\pm$0.024 & 0.677$\pm$0.039$\pm$0.019 & 0.633$\pm$0.058$\pm$0.018 \\
 & $2.36-2.75$ & 0.811$\pm$0.039$\pm$0.018 & 0.634$\pm$0.031$\pm$0.014 & 0.554$\pm$0.045$\pm$0.013 \\
 & $2.75-3.14$ & 0.811$\pm$0.038$\pm$0.020 & 0.652$\pm$0.031$\pm$0.017 & 0.569$\pm$0.044$\pm$0.017 \\
\hline

    \end{tabular}
    \caption{Same as Table~\ref{tab:all_R_data}, except in slices of $\Delta Y$.  These values are shown in the bottom row of Fig.~\ref{fig:dY}.}
    \label{tab:dy_R_data}
\end{table*}

\begin{table*}[]
    \centering
    \begin{tabular}{c|c|c|c|c}
        $p^T_1$ [GeV] & $\Delta\phi$ [rad] & C & Fe & Pb \\
         \hline
         $0.25-0.40$ & $0.00-0.39$ & 1.094$\pm$0.095$\pm$0.035 & 1.077$\pm$0.088$\pm$0.041 & 1.248$\pm$0.146$\pm$0.056 \\
 & $0.39-0.79$ & 1.003$\pm$0.081$\pm$0.027 & 0.978$\pm$0.075$\pm$0.027 & 0.942$\pm$0.113$\pm$0.031 \\
 & $0.79-1.18$ & 1.064$\pm$0.078$\pm$0.024 & 0.894$\pm$0.065$\pm$0.021 & 1.000$\pm$0.108$\pm$0.024 \\
 & $1.18-1.57$ & 1.056$\pm$0.067$\pm$0.031 & 0.837$\pm$0.054$\pm$0.025 & 0.902$\pm$0.088$\pm$0.027 \\
 & $1.57-1.96$ & 0.867$\pm$0.046$\pm$0.021 & 0.731$\pm$0.039$\pm$0.018 & 0.755$\pm$0.063$\pm$0.019 \\
 & $1.96-2.36$ & 0.862$\pm$0.044$\pm$0.021 & 0.722$\pm$0.037$\pm$0.018 & 0.759$\pm$0.060$\pm$0.019 \\
 & $2.36-2.75$ & 0.830$\pm$0.041$\pm$0.021 & 0.726$\pm$0.035$\pm$0.018 & 0.570$\pm$0.049$\pm$0.015 \\
 & $2.75-3.14$ & 0.879$\pm$0.042$\pm$0.020 & 0.765$\pm$0.036$\pm$0.019 & 0.628$\pm$0.051$\pm$0.017 \\
\hline
$0.40-0.60$ & $0.00-0.39$ & 1.517$\pm$0.161$\pm$0.095 & 1.611$\pm$0.158$\pm$0.130 & 1.624$\pm$0.229$\pm$0.179 \\
 & $0.39-0.79$ & 1.577$\pm$0.145$\pm$0.048 & 1.386$\pm$0.125$\pm$0.048 & 1.354$\pm$0.180$\pm$0.057 \\
 & $0.79-1.18$ & 1.355$\pm$0.104$\pm$0.033 & 1.044$\pm$0.082$\pm$0.025 & 1.232$\pm$0.136$\pm$0.030 \\
 & $1.18-1.57$ & 1.115$\pm$0.066$\pm$0.026 & 0.939$\pm$0.055$\pm$0.022 & 1.060$\pm$0.090$\pm$0.026 \\
 & $1.57-1.96$ & 0.952$\pm$0.045$\pm$0.023 & 0.781$\pm$0.037$\pm$0.019 & 0.945$\pm$0.064$\pm$0.023 \\
 & $1.96-2.36$ & 0.844$\pm$0.035$\pm$0.022 & 0.673$\pm$0.028$\pm$0.017 & 0.628$\pm$0.042$\pm$0.016 \\
 & $2.36-2.75$ & 0.763$\pm$0.027$\pm$0.016 & 0.616$\pm$0.022$\pm$0.013 & 0.580$\pm$0.034$\pm$0.013 \\
 & $2.75-3.14$ & 0.778$\pm$0.026$\pm$0.018 & 0.618$\pm$0.021$\pm$0.014 & 0.598$\pm$0.032$\pm$0.015 \\
\hline
$0.60-1.00$ & $0.00-0.39$ & 3.201$\pm$0.659$\pm$0.545 & 3.793$\pm$0.726$\pm$0.734 & 3.053$\pm$0.780$\pm$0.661 \\
 & $0.39-0.79$ & 3.058$\pm$0.564$\pm$0.390 & 3.374$\pm$0.585$\pm$0.431 & 2.802$\pm$0.652$\pm$0.358 \\
 & $0.79-1.18$ & 2.067$\pm$0.274$\pm$0.048 & 2.252$\pm$0.277$\pm$0.053 & 2.470$\pm$0.392$\pm$0.063 \\
 & $1.18-1.57$ & 1.494$\pm$0.128$\pm$0.045 & 1.333$\pm$0.110$\pm$0.040 & 1.277$\pm$0.152$\pm$0.038 \\
 & $1.57-1.96$ & 1.376$\pm$0.098$\pm$0.049 & 1.193$\pm$0.083$\pm$0.043 & 1.215$\pm$0.119$\pm$0.043 \\
 & $1.96-2.36$ & 0.996$\pm$0.054$\pm$0.022 & 0.809$\pm$0.043$\pm$0.018 & 0.761$\pm$0.062$\pm$0.017 \\
 & $2.36-2.75$ & 0.860$\pm$0.033$\pm$0.019 & 0.629$\pm$0.025$\pm$0.014 & 0.645$\pm$0.038$\pm$0.015 \\
 & $2.75-3.14$ & 0.676$\pm$0.024$\pm$0.015 & 0.549$\pm$0.019$\pm$0.013 & 0.507$\pm$0.029$\pm$0.015 \\
\hline

    \end{tabular}
    \caption{Same as Table~\ref{tab:all_R_data}, except in slices of $p^T_1$. These values are shown in the bottom row of Fig.~\ref{fig:pT1}.}
    \label{tab:pT1_R_data}
\end{table*}

\begin{table*}[]
    \centering
         
    \begin{tabular}{c|c|c|c|c}
         $p^T_1$ [GeV]& $\Delta\phi$ [rad] & C & Fe & Pb \\
         \hline
         $0.25-0.40$ & $0.00-0.39$ & 1.044$\pm$0.081$\pm$0.045 & 1.189$\pm$0.083$\pm$0.076 & 1.203$\pm$0.122$\pm$0.109 \\
 & $0.39-0.79$ & 1.058$\pm$0.073$\pm$0.024 & 1.037$\pm$0.067$\pm$0.027 & 0.959$\pm$0.096$\pm$0.032 \\
 & $0.79-1.18$ & 1.169$\pm$0.071$\pm$0.024 & 0.991$\pm$0.059$\pm$0.021 & 1.041$\pm$0.091$\pm$0.022 \\
 & $1.18-1.57$ & 1.027$\pm$0.049$\pm$0.022 & 0.884$\pm$0.042$\pm$0.019 & 0.926$\pm$0.065$\pm$0.022 \\
 & $1.57-1.96$ & 0.858$\pm$0.035$\pm$0.018 & 0.733$\pm$0.030$\pm$0.015 & 0.812$\pm$0.048$\pm$0.018 \\
 & $1.96-2.36$ & 0.799$\pm$0.030$\pm$0.017 & 0.650$\pm$0.024$\pm$0.014 & 0.636$\pm$0.037$\pm$0.014 \\
 & $2.36-2.75$ & 0.753$\pm$0.025$\pm$0.016 & 0.617$\pm$0.021$\pm$0.014 & 0.601$\pm$0.032$\pm$0.015 \\
 & $2.75-3.14$ & 0.705$\pm$0.023$\pm$0.016 & 0.598$\pm$0.019$\pm$0.018 & 0.555$\pm$0.029$\pm$0.025 \\
\hline
$0.40-0.60$ & $0.00-0.39$ & 2.051$\pm$0.231$\pm$0.169 & 1.705$\pm$0.190$\pm$0.203 & 1.642$\pm$0.262$\pm$0.243 \\
 & $0.39-0.79$ & 1.748$\pm$0.185$\pm$0.079 & 1.656$\pm$0.168$\pm$0.101 & 1.362$\pm$0.211$\pm$0.098 \\
 & $0.79-1.18$ & 1.296$\pm$0.110$\pm$0.028 & 1.136$\pm$0.094$\pm$0.025 & 1.436$\pm$0.160$\pm$0.036 \\
 & $1.18-1.57$ & 1.286$\pm$0.086$\pm$0.027 & 1.030$\pm$0.069$\pm$0.022 & 1.077$\pm$0.106$\pm$0.023 \\
 & $1.57-1.96$ & 1.059$\pm$0.055$\pm$0.023 & 0.871$\pm$0.045$\pm$0.019 & 0.887$\pm$0.069$\pm$0.019 \\
 & $1.96-2.36$ & 0.936$\pm$0.041$\pm$0.020 & 0.726$\pm$0.032$\pm$0.016 & 0.695$\pm$0.048$\pm$0.015 \\
 & $2.36-2.75$ & 0.863$\pm$0.030$\pm$0.018 & 0.670$\pm$0.024$\pm$0.014 & 0.584$\pm$0.034$\pm$0.013 \\
 & $2.75-3.14$ & 0.799$\pm$0.026$\pm$0.017 & 0.627$\pm$0.020$\pm$0.014 & 0.575$\pm$0.030$\pm$0.013 \\
\hline
$0.60-0.80$ & $0.00-0.39$ & 2.054$\pm$0.975$\pm$0.051 & 2.889$\pm$1.205$\pm$0.071 & 3.626$\pm$1.866$\pm$0.089 \\
 & $0.39-0.79$ & 5.800$\pm$2.387$\pm$0.191 & 3.925$\pm$1.660$\pm$0.129 & 5.747$\pm$2.806$\pm$0.189 \\
 & $0.79-1.18$ & 3.313$\pm$1.053$\pm$0.108 & 1.389$\pm$0.518$\pm$0.046 & 2.373$\pm$1.050$\pm$0.077 \\
 & $1.18-1.57$ & 2.655$\pm$0.682$\pm$0.059 & 2.340$\pm$0.589$\pm$0.052 & 2.419$\pm$0.819$\pm$0.053 \\
 & $1.57-1.96$ & 2.401$\pm$0.423$\pm$0.056 & 1.654$\pm$0.303$\pm$0.039 & 2.511$\pm$0.561$\pm$0.059 \\
 & $1.96-2.36$ & 1.337$\pm$0.150$\pm$0.029 & 1.119$\pm$0.124$\pm$0.024 & 1.014$\pm$0.173$\pm$0.022 \\
 & $2.36-2.75$ & 0.948$\pm$0.068$\pm$0.021 & 0.715$\pm$0.053$\pm$0.015 & 0.681$\pm$0.079$\pm$0.015 \\
 & $2.75-3.14$ & 0.857$\pm$0.057$\pm$0.019 & 0.686$\pm$0.046$\pm$0.015 & 0.663$\pm$0.069$\pm$0.014 \\
\hline

    \end{tabular}
    \caption{Same as Table~\ref{tab:all_R_data}, except in slices of $p^T_2$. These values are shown in the bottom row of Fig.~\ref{fig:pT2}.}
    \label{tab:pT2_R_data}
\end{table*}

\begin{table*}[]
    \centering
    \begin{tabular}{c|c|c|c|c}
    Slice & D & C & Fe & Pb\\
         \hline     Integrated result& 1.114$\pm$0.005$\pm$0.030& 1.240$\pm$0.008$\pm$0.034& 1.285$\pm$0.008$\pm$0.036& 1.313$\pm$0.014$\pm$0.038\\
\hline
-0.50 < $\Delta Y$< 0.50& 0.877$\pm$0.013$\pm$0.029& 1.030$\pm$0.021$\pm$0.033& 1.087$\pm$0.022$\pm$0.035& 1.117$\pm$0.035$\pm$0.035\\
0.50 < $\Delta Y$< 1.50& 1.081$\pm$0.006$\pm$0.027& 1.229$\pm$0.010$\pm$0.037& 1.274$\pm$0.010$\pm$0.043& 1.275$\pm$0.017$\pm$0.049\\
1.50 < $\Delta Y$< 2.50& 1.294$\pm$0.011$\pm$0.037& 1.365$\pm$0.016$\pm$0.039& 1.410$\pm$0.017$\pm$0.041& 1.499$\pm$0.028$\pm$0.043\\
\hline
0.25 < $p^T_{1}$< 0.40 GeV& 1.375$\pm$0.009$\pm$0.031& 1.440$\pm$0.015$\pm$0.032& 1.464$\pm$0.015$\pm$0.033& 1.539$\pm$0.026$\pm$0.035\\
0.40 < $p^T_{1}$< 0.60 GeV& 1.076$\pm$0.007$\pm$0.021& 1.230$\pm$0.012$\pm$0.025& 1.264$\pm$0.012$\pm$0.028& 1.298$\pm$0.021$\pm$0.030\\
0.60 < $p^T_{1}$< 1.00 GeV& 0.808$\pm$0.008$\pm$0.020& 1.037$\pm$0.014$\pm$0.024& 1.118$\pm$0.014$\pm$0.028& 1.107$\pm$0.024$\pm$0.026\\
\hline
0.25 < $p^T_{2}$< 0.40 GeV& 1.220$\pm$0.007$\pm$0.019& 1.332$\pm$0.011$\pm$0.021& 1.383$\pm$0.011$\pm$0.023& 1.398$\pm$0.018$\pm$0.023\\
0.40 < $p^T_{2}$< 0.60 GeV& 1.002$\pm$0.008$\pm$0.018& 1.169$\pm$0.013$\pm$0.026& 1.199$\pm$0.013$\pm$0.030& 1.233$\pm$0.022$\pm$0.032\\
0.60 < $p^T_{2}$< 0.80 GeV& 0.699$\pm$0.017$\pm$0.018& 0.939$\pm$0.027$\pm$0.018& 0.936$\pm$0.029$\pm$0.019& 1.025$\pm$0.051$\pm$0.019\\
\hline

    \end{tabular}
    \caption{Tabulation of the widths for the integrated measurement (top row) and sliced measurements.  These correspond to the top panels of Figs.~\ref{fig:widths_1d}, \ref{fig:dY_summary}, \ref{fig:pT1_summary}, and \ref{fig:pT2_summary}.}
    \label{tab:width_data}
\end{table*}
\begin{table*}[]
    \centering
    \begin{tabular}{c|c|c|c}
    Slice & C & Fe & Pb\\
         \hline
         Integrated result& 0.544$\pm$0.021$\pm$0.022& 0.641$\pm$0.018$\pm$0.030& 0.695$\pm$0.027$\pm$0.033\\
\hline
-0.50 < $\Delta Y$< 0.50& 0.540$\pm$0.045$\pm$0.015& 0.643$\pm$0.040$\pm$0.023& 0.693$\pm$0.058$\pm$0.020\\
0.50 < $\Delta Y$< 1.50& 0.585$\pm$0.023$\pm$0.039& 0.674$\pm$0.021$\pm$0.051& 0.677$\pm$0.033$\pm$0.065\\
1.50 < $\Delta Y$< 2.50& 0.432$\pm$0.062$\pm$0.020& 0.560$\pm$0.049$\pm$0.026& 0.756$\pm$0.059$\pm$0.034\\
\hline
0.25 < $p^T_{1}$< 0.40 GeV& 0.431$\pm$0.058$\pm$0.014& 0.504$\pm$0.050$\pm$0.017& 0.693$\pm$0.060$\pm$0.024\\
0.40 < $p^T_{1}$< 0.60 GeV& 0.595$\pm$0.028$\pm$0.021& 0.663$\pm$0.026$\pm$0.030& 0.725$\pm$0.038$\pm$0.033\\
0.60 < $p^T_{1}$< 1.00 GeV& 0.650$\pm$0.024$\pm$0.021& 0.772$\pm$0.022$\pm$0.028& 0.756$\pm$0.036$\pm$0.025\\
\hline
0.25 < $p^T_{2}$< 0.40 GeV& 0.534$\pm$0.031$\pm$0.016& 0.652$\pm$0.026$\pm$0.023& 0.682$\pm$0.039$\pm$0.025\\
0.40 < $p^T_{2}$< 0.60 GeV& 0.603$\pm$0.027$\pm$0.032& 0.659$\pm$0.026$\pm$0.040& 0.718$\pm$0.039$\pm$0.041\\
0.60 < $p^T_{2}$< 0.80 GeV& 0.627$\pm$0.045$\pm$0.013& 0.622$\pm$0.047$\pm$0.013& 0.750$\pm$0.070$\pm$0.016\\
\hline

    \end{tabular}
    \caption{Tabulation of the broadenings for the integrated measurement (top row) and sliced measurements. These correspond to the bottom panels of Figs.~\ref{fig:widths_1d}, \ref{fig:dY_summary}, \ref{fig:pT1_summary}, and \ref{fig:pT2_summary}.}
    \label{tab:broadening_data}
\end{table*}
\FloatBarrier

\section{Multi-Dimensional Correlation Predictions in the eHIJING and BeAGLE Event Generators}\label{app:more_models}
\begin{figure}[b]
    \centering
    \includegraphics[width=0.5\textwidth]{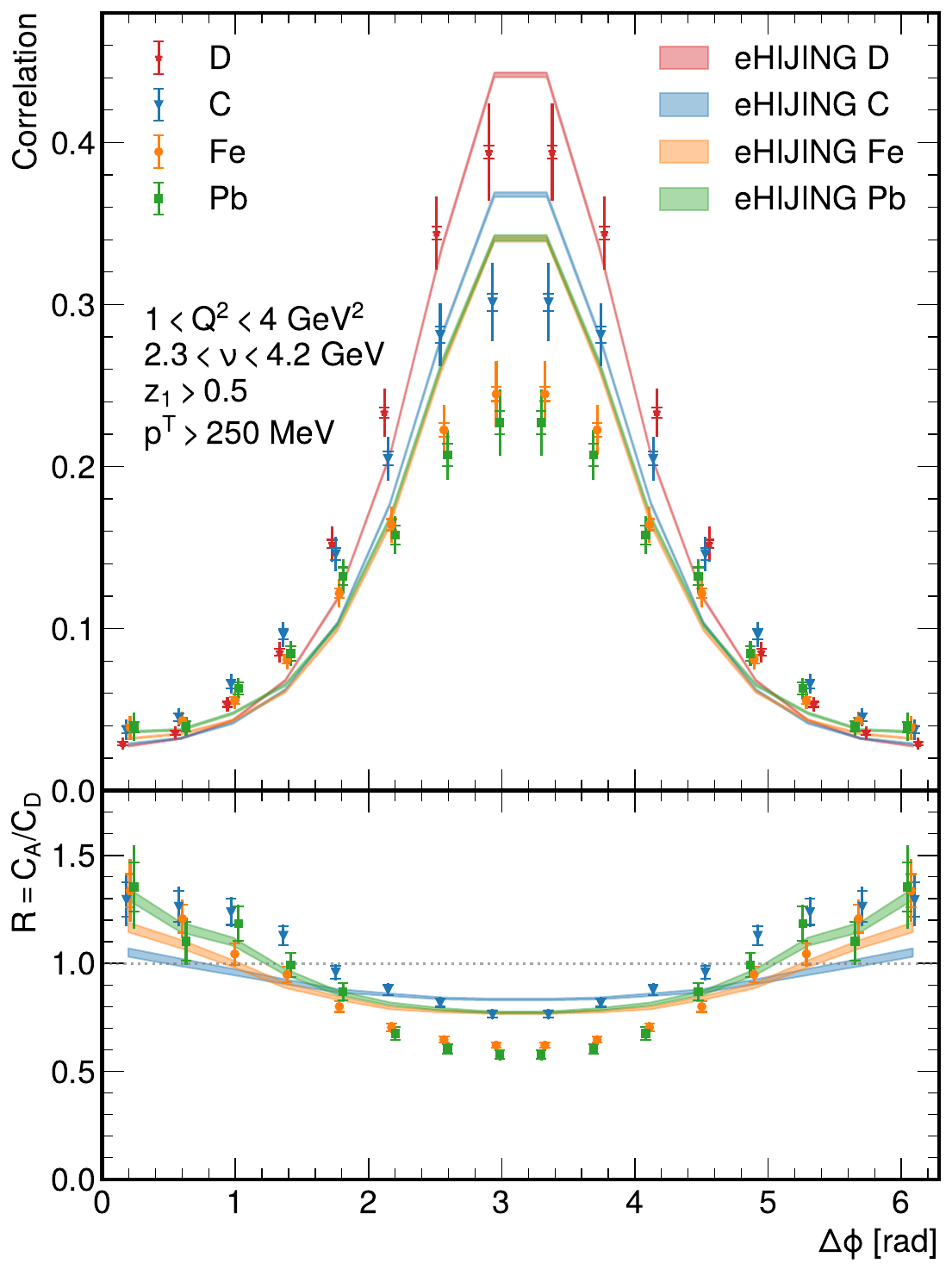}
    \caption{Comparison of our measured correlation functions to those calculated by the eHIJING model.}
    \label{fig:eHIJING_integrated}
\end{figure}

\begin{figure}[b]
    \centering
    \includegraphics[width=0.5\textwidth]{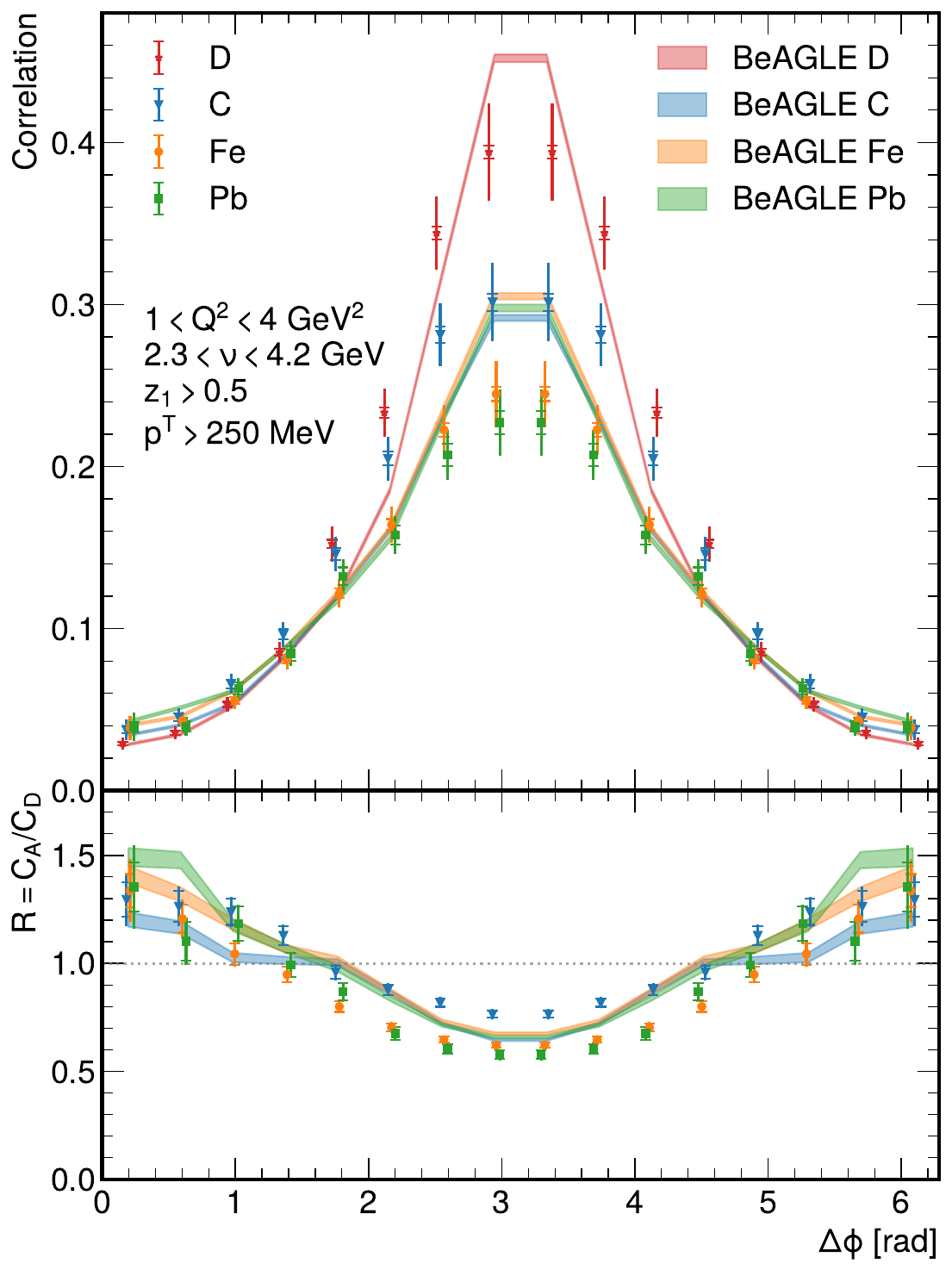}
    \caption{Comparison of our measured correlation functions to those calculated by the BeAGLE event generator.}
    \label{fig:BeAGLE_integrated}
\end{figure}

\begin{figure*}
    \centering
    \includegraphics[width=\textwidth]{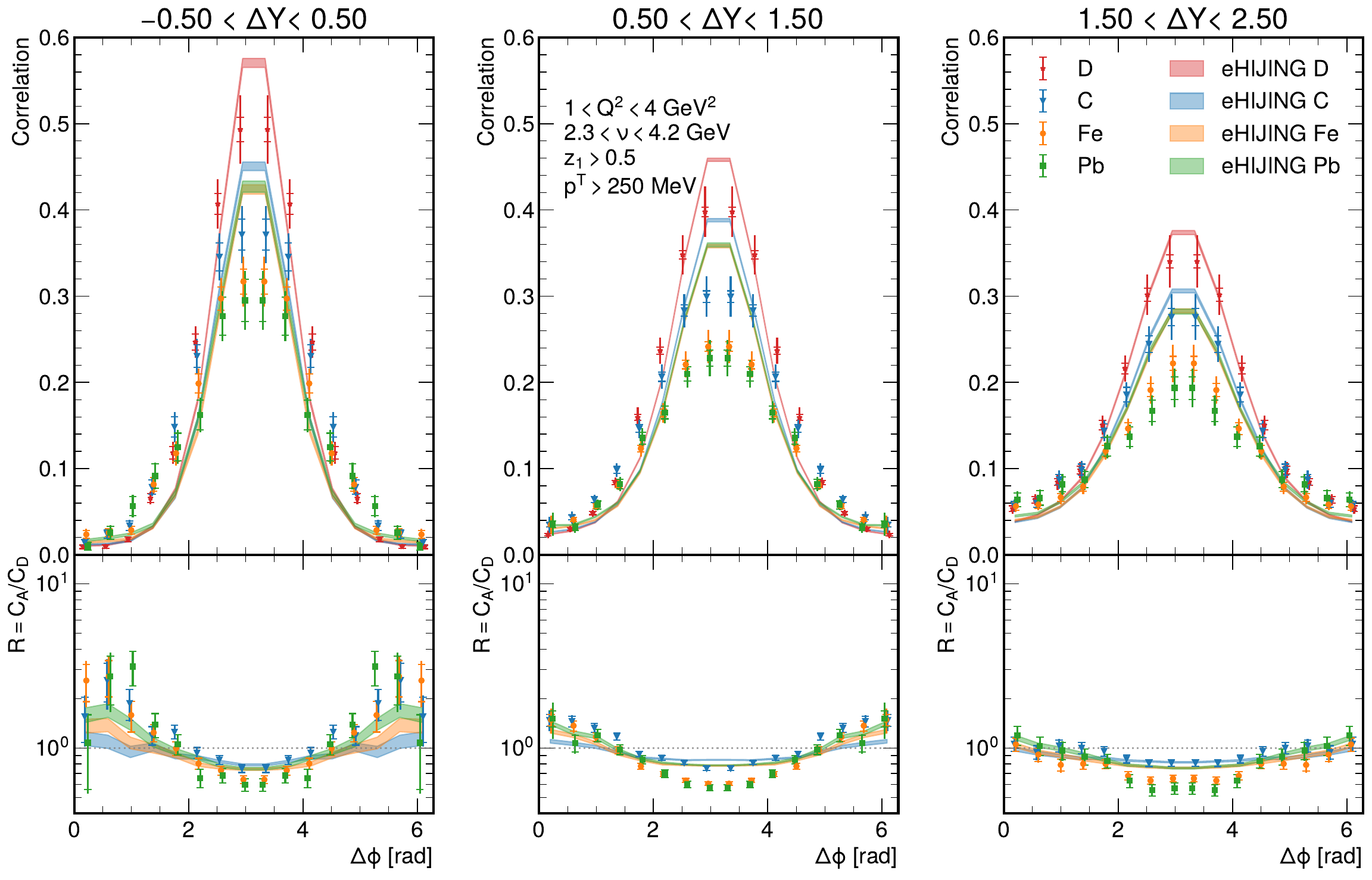}
    \caption{Comparison of our measured correlation functions in slices of $\Delta Y$ to those calculated by the eHIJING model.}
    \label{fig:eHIJING_dY}
\end{figure*}

\begin{figure*}
    \centering
    \includegraphics[width=\textwidth]{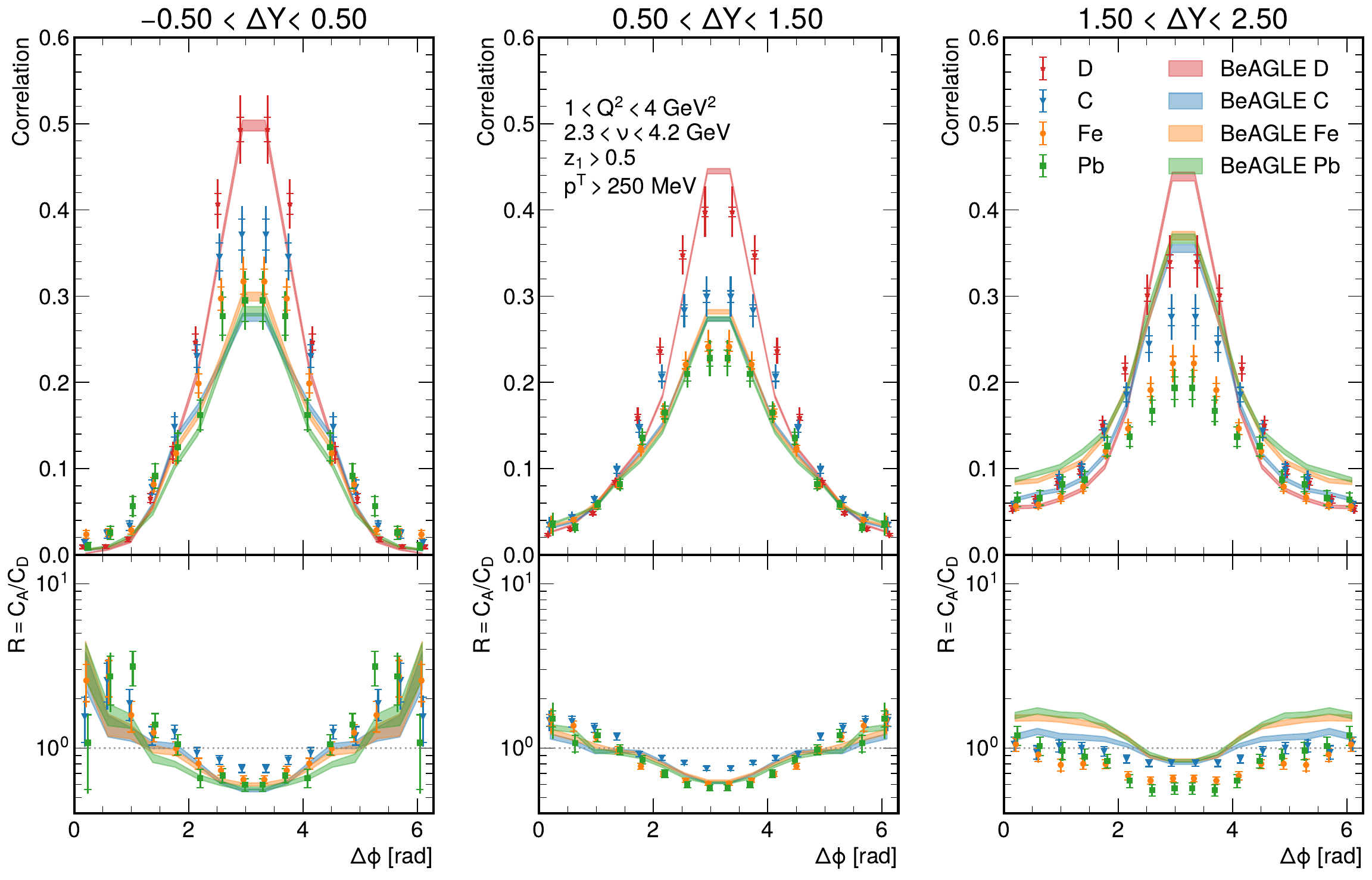}
    \caption{Comparison of our measured correlation functions in slices of $\Delta Y$ to those calculated by the BeAGLE event generator.}
    \label{fig:BeAGLE_dY}
\end{figure*}

\begin{figure*}
    \centering
    \includegraphics[width=\textwidth]{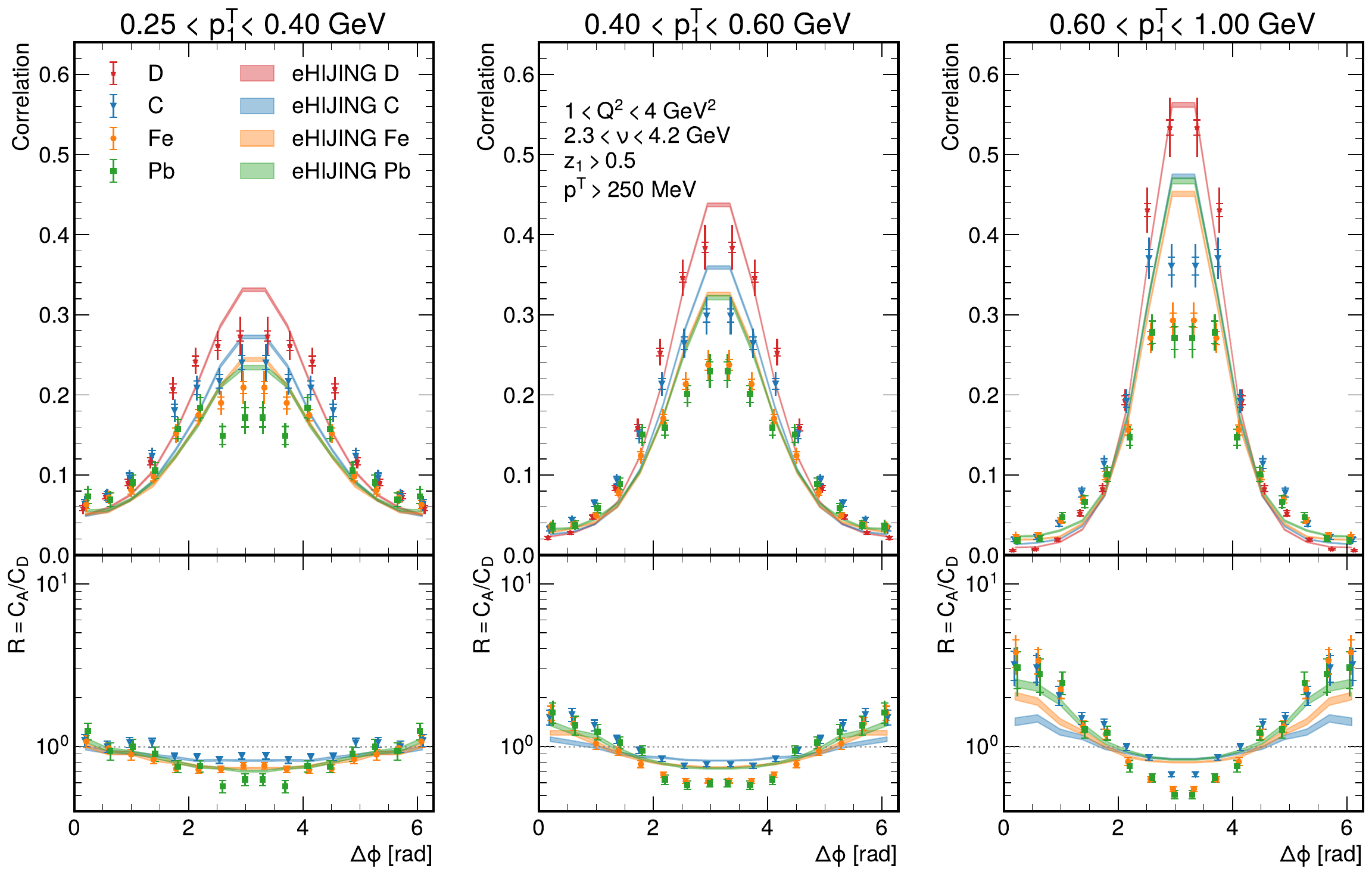}
    \caption{Comparison of our measured correlation functions in slices of $p^T_1$ to those calculated by the eHIJING model.}
    \label{fig:eHIJING_pT1}
\end{figure*}

\begin{figure*}
    \centering
    \includegraphics[width=\textwidth]{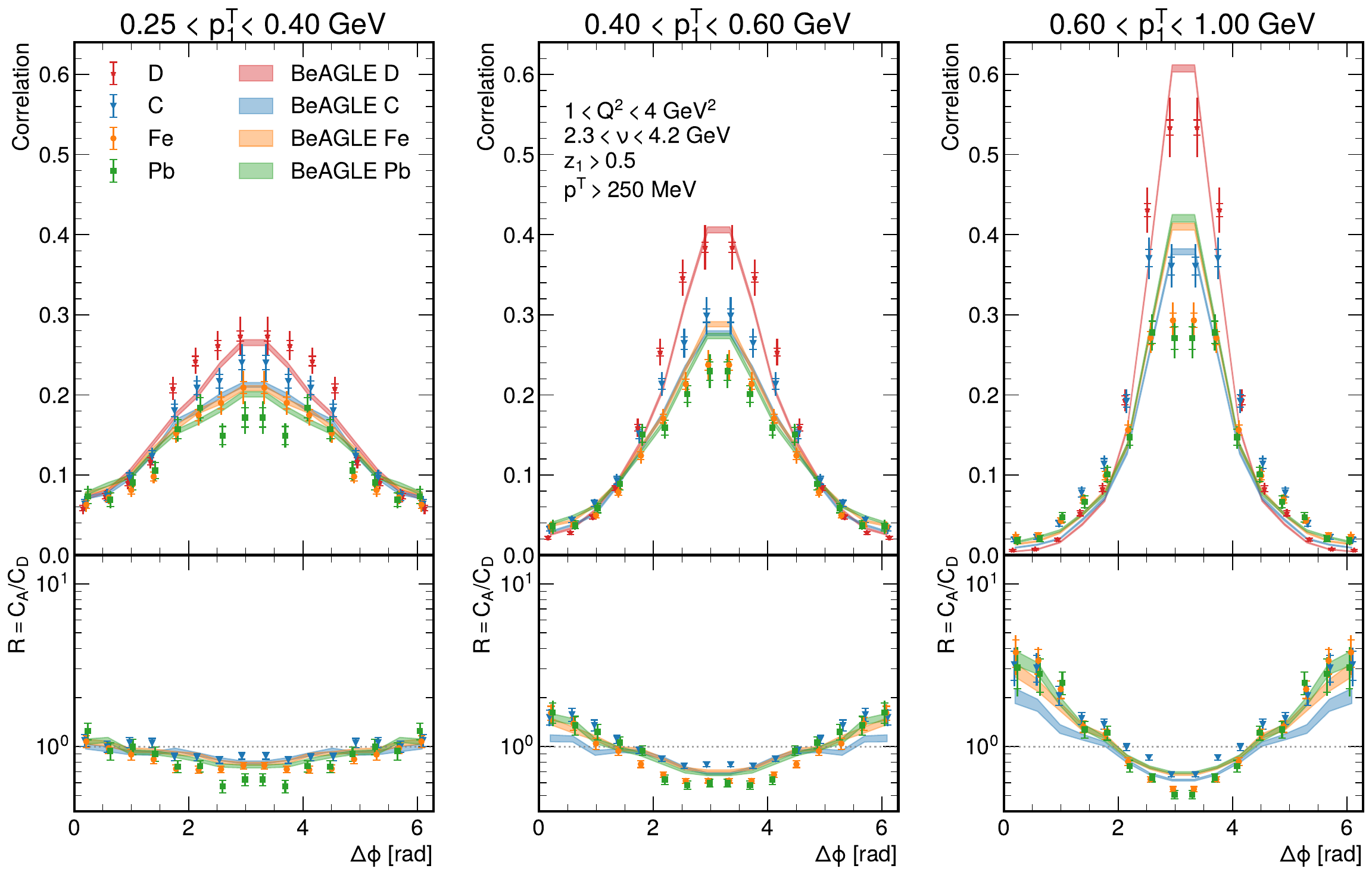}
    \caption{Comparison of our measured correlation functions in slices of $p^T_1$ to those calculated by the BeAGLE event generator.}
    \label{fig:BeAGLE_pT1}
\end{figure*}

\begin{figure*}
    \centering
    \includegraphics[width=\textwidth]{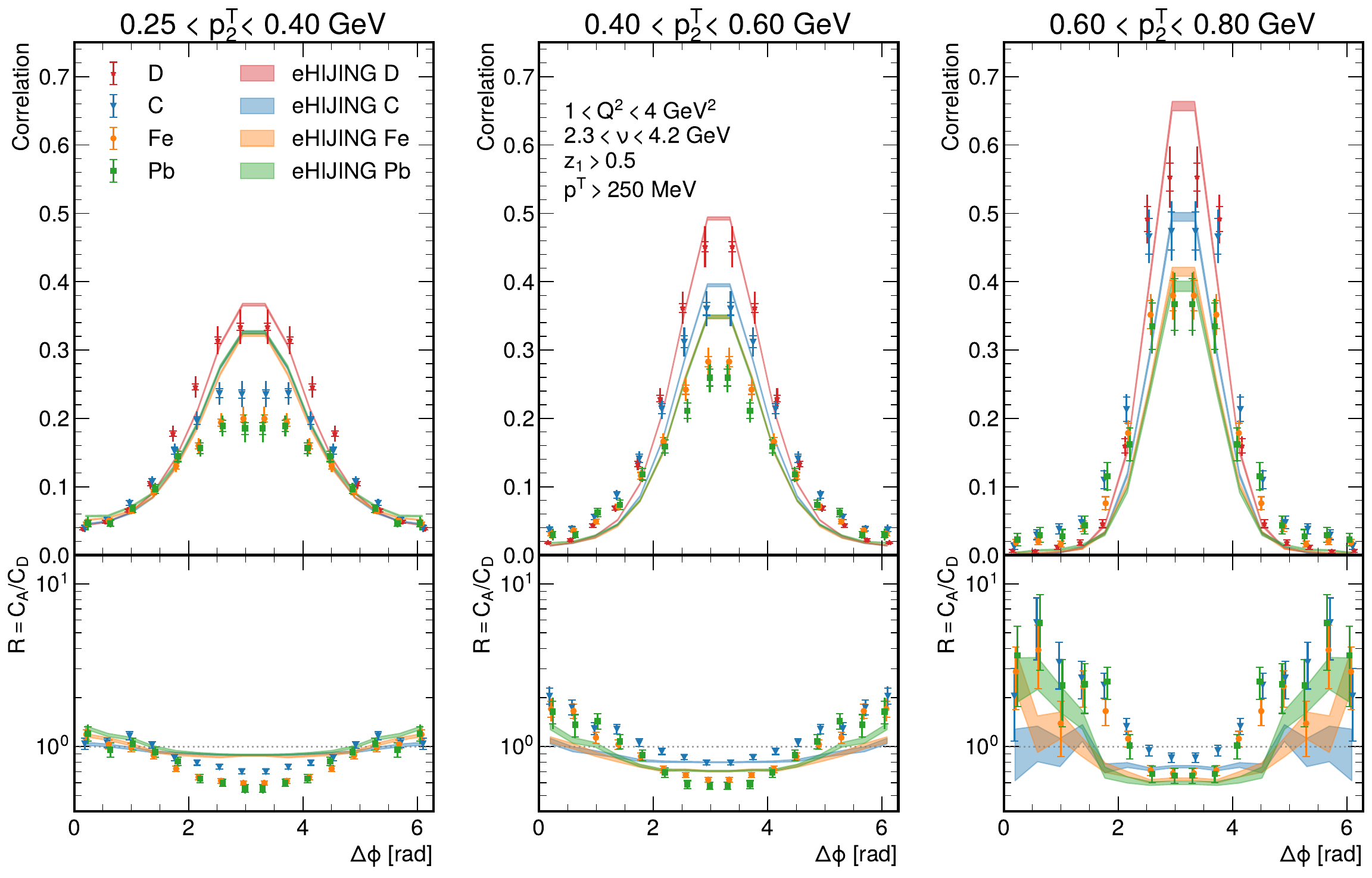}
    \caption{Comparison of our measured correlation functions in slices of $p^T_2$ to those calculated by the eHIJING model.}
    \label{fig:eHIJING_pT2}
\end{figure*}

\begin{figure*}
    \centering
    \includegraphics[width=\textwidth]{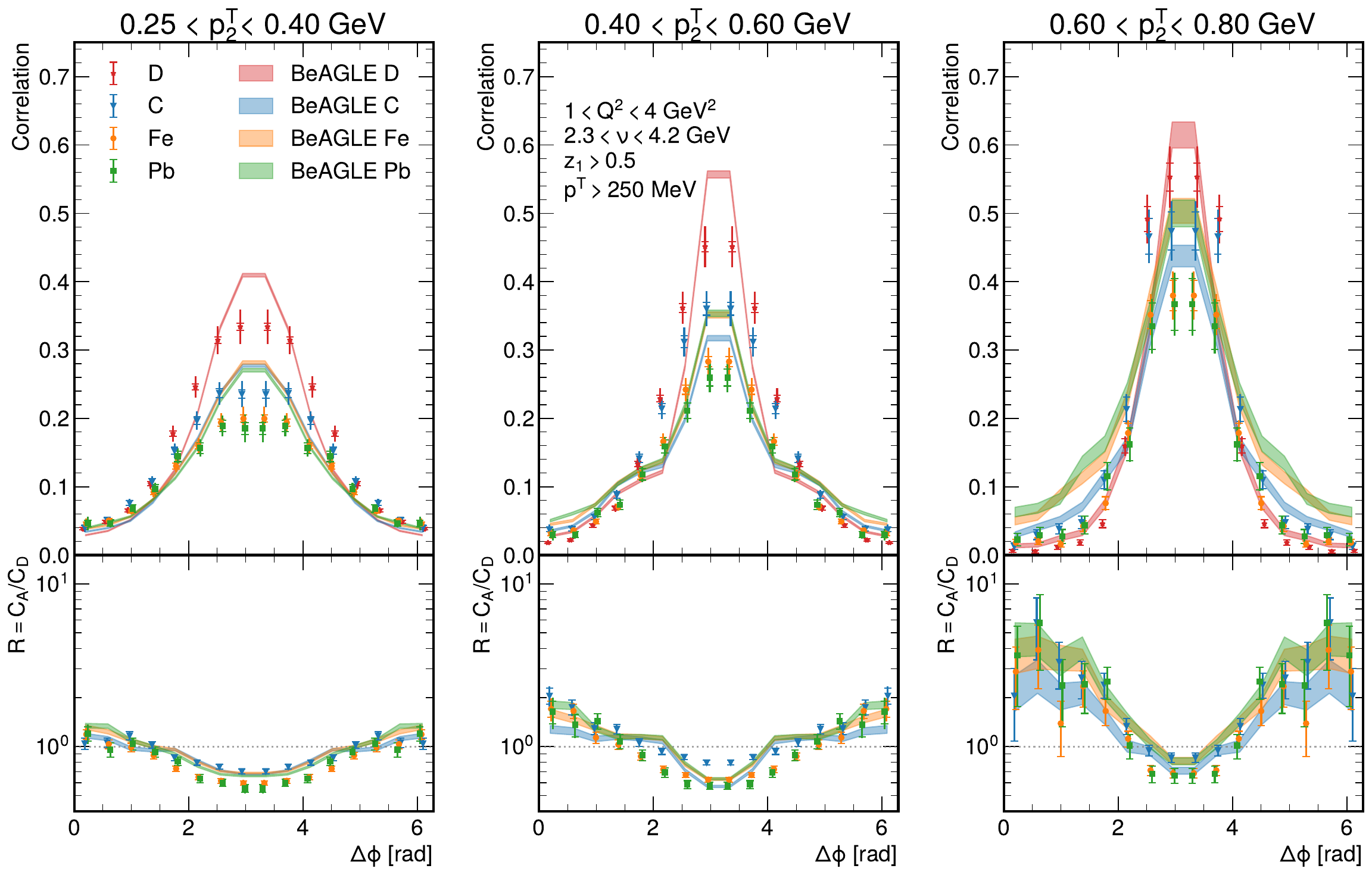}
    \caption{Comparison of our measured correlation functions in slices of $p^T_2$ to those calculated by the BeAGLE event generator.}
    \label{fig:BeAGLE_pT2}
\end{figure*}

{In Figs.~\ref{fig:eHIJING_integrated} and \ref{fig:BeAGLE_integrated}, we compare our measured correlation functions to those of the eHIJING and BeAGLE event generators.  Likewise, we compare our data to these models' predictions in slices of $\Delta Y$ in Figs.~\ref{fig:eHIJING_dY} and \ref{fig:BeAGLE_dY}, in slices of $p^T_1$ in Figs.~\ref{fig:eHIJING_pT1} and \ref{fig:BeAGLE_pT1}, and in slices of $p^T_2$ in Figs.~\ref{fig:eHIJING_pT2} and \ref{fig:BeAGLE_pT2}.}

\FloatBarrier

\end{document}